\documentclass{pasj00}

\newcommand{\kt}{\ensuremath{k_{\rm{B}}T}}
\newcommand{\lx}{\ensuremath{L_{\rm{X}}}}

\newcommand{\fx}{\ensuremath{F_{\rm{X}}}}
\newcommand{\nh}{\ensuremath{N_{\rm{H}}}}

\SetRunningHead{D.~Takei et al.}{X-ray Study of Rekindled Accretion in the Classical Nova V2491 Cygni}

\Received{2010/11/13} 
\Accepted{2011/02/01} 
\Published{}

\begin{document}

\title{X-ray Study of Rekindled Accretion in the Classical Nova V2491 Cygni}
\author{
Dai~\textsc{Takei},\altaffilmark{1}
Jan-Uwe~\textsc{Ness},\altaffilmark{2}
Masahiro~\textsc{Tsujimoto},\altaffilmark{3}
Shunji~\textsc{Kitamoto},\altaffilmark{1}
Jeremy~J.~\textsc{Drake},\altaffilmark{4}
\\
Julian~P.~\textsc{Osborne},\altaffilmark{5}
Hiromitsu~\textsc{Takahashi},\altaffilmark{6}
and Kenzo~\textsc{Kinugasa}\altaffilmark{7}
}
\altaffiltext{1}{Department of Physics, Rikkyo University, 3-34-1 Nishi-Ikebukuro, Toshima, Tokyo 171-8501}
\email{takei@ast.rikkyo.ac.jp}
\altaffiltext{2}{European Space Agency, XMM-Newton Observatory SOC, SRE-OAX, \\
                 Apartado 78, 28691 Villanueva de la Ca\~nada, Madrid, Spain}
\altaffiltext{3}{Japan Aerospace Exploration Agency, Institute of Space and Astronautical Science, \\
                 3-1-1 Yoshino-dai, Chuo-ku, Sagamihara, Kanagawa 252-5210}
\altaffiltext{4}{Smithsonian Astrophysical Observatory, MS-3, 60 Garden Street, Cambridge, MA 02138, USA}
\altaffiltext{5}{Department of Physics \& Astronomy, University of Leicester, Leicester, LE8 9DX, UK}
\altaffiltext{6}{Hiroshima Astrophysical Science Center, Hiroshima University, \\
                 1-3-1 Kagamiyama, Higashi-Hiroshima, Hiroshima 739-8526}
\altaffiltext{7}{Gunma Astronomical Observatory, 6860-86 Nakayama, Takayama, Agatsuma, Gunma 377-0702}
\maketitle

\begin{abstract}
 We conducted an X-ray spectroscopic study of the classical nova V2491 Cygni using our
 target-of-opportunity observation data with the Suzaku and XMM-Newton satellites as
 well as archived data with the Swift satellite. Medium-resolution ($R$ $\sim$ 10--50)
 spectra were obtained using the X-ray CCD spectrometers at several post-nova epochs on
 days 9, 29, 40, 50, and 60--150 in addition to a pre-nova interval between days $-$322
 and $-$100 all relative to the time when the classical nova was spotted. We found
 remarkable changes in the time series of the spectra: (a) In the pre-nova phase and on
 day 9, the 6.7~keV emission line from Fe\emissiontype{XXV} was significantly detected.
 (b) On day 29, no such emission line was found. (c) On day 40, the 6.7~keV emission
 line emerged again. (d) On days 50 and 60--150, three emission lines at 6.4, 6.7, and
 7.0~keV respectively from quasi-neutral Fe, Fe\emissiontype{XXV}, and
 Fe\emissiontype{XXVI} were found. Statistically significant changes of the Fe~K line
 intensities were confirmed between day 29 and 50. Based on these phenomena, we conclude
 that (1) the post-nova evolution can be divided into two different phases, (2) ejecta
 is responsible for the X-ray emission in the earlier phase, while rekindled accretion
 is for the later phase, and (3) the accretion process is considered to be reestablished
 as early as day 50 when the quasi-neutral Fe emission line emerged, which is a common
 signature of accretion from magnetic cataclysmic variables.
\end{abstract}

\section{Introduction}\label{introduction}
A classical nova is an outburst characterized by a sudden increase in optical brightness
by $\sim$10~mag. It is observed in binary systems composed of a white dwarf (WD) and a
main-sequence or a giant star. Hydrogen-rich material filling the Roche-lobe around the
companion accretes onto the WD. When the amount of accreted material reaches a critical
mass, hydrogen fusion is ignited explosively, causing a thermonuclear runaway on the WD
surface (e.g., \cite{starrfield2008}). The sudden increase in radiation pressure leads
to the ejection of accreted material, and a rapid and large rise in the optical
brightness is observed as a classical nova. For reviews of classical novae, readers can
refer to e.g., \citet{warner2003} and \citet{bode2008}.

The form of mass accretion mainly depends on the strength of the magnetic field on the
WD ($B_{\rm{WD}}$; e.g., \cite{warner2003}). For non-magnetic or weakly magnetized WDs
($B_{\rm{WD}}$ $\lesssim$ 10$^{5}$ gauss), the accretion disk is formed in the binary
system. For strongly magnetized WDs, called polars ($B_{\rm{WD}}$ $\gtrsim$ 10$^{7}$
gauss), the motion of infalling material is governed by their strong magnetic field,
thus instead of a disk an accretion funnel is formed along the magnetic field lines
(e.g., \cite{cropper1990}). For those with magnetic field strengths between these two,
called intermediate polars (IPs), the accretion disk is truncated at the Alfv\'{e}n
radius where the magnetic and the ram pressures are equal to each other, from which
matter accretes along the field lines (e.g., \cite{patterson1994}). In principle,
classical nova explosions can occur regardless of the strength of $B_{\rm{WD}}$, but
most observed cases are from non-magnetic or weakly magnetized WDs. Classical novae from
magnetized WDs so far make up only a handful of cases (e.g., V1500 Cyg, GK Per, and CP
Pup; \cite{warner2003,warner2008}).

During a nova eruption, the accretion process is considered to be suspended until the
radiation pressure drops at the end of nuclear processing on the WD surface (e.g.,
\cite{prialnik1986a,prialnik1986b,shara1986}). Following this, a reestablished accretion
process can start to accumulate material again for the next eruption. This typically
takes a long time ($\sim$10$^{4-5}$~yr), except for a handful of ``recurrent'' novae
that undergo outbursts every $\sim$10--100~yr (e.g., \cite{schaefer2010a}). It is
currently unknown how early the accretion process resumes after a nova
eruption. \citet{drake2010} showed that an accretion disk can be completely destroyed in
the blast. At least for the shorter period systems, a disk is expected to be rejuvenated
for IPs and weakly magnetized WDs prior to accretion resumption onto the WD. The
reestablishment of accretion is a key event, which affects the post-nova development and
the ultimate fate of the binary system (e.g., \cite{prialnik1982,prialnik1995}).
However, observations to probe rekindled accretion are difficult in the immediate
aftermath of a nova, as thick material still remains around the binary system as a
consequence of envelope ejection.

Moderately hard X-ray (2.0--10~keV) observations of classical novae in magnetized WDs
are one of the few tools to enable direct access to rekindled accretion. X-rays can
penetrate through the thick circumbinary material, allowing us to delve into the system
immediately after the nova eruption. In particular, IPs emit bright X-rays of
$\sim$10$^{33}$~erg~s$^{-1}$ in the 2.0--10~keV energy band (e.g.,
\cite{chanmugam1991,norton1989,warner2003}), which originate from high temperature
plasma produced by the accretion shock on the WD surface. Several observational methods
in the moderately hard X-ray band can be used to probe the accretion process after
classical novae: (1) a quasi-neutral Fe (hereafter called Fe\emissiontype{I}) K$\alpha$
emission line at 6.4~keV, which is produced as fluorescence at the WD surface
illuminated by hard X-ray photons of shock-excited plasma at the base of the accretion
column (e.g., \cite{ezuka1999}), (2) flickering in the moderately hard X-ray light curve
on a time scale of seconds to minutes caused by random processes in accretion (e.g.,
\cite{ribeiro2008}), and (3) X-ray periodicity due to a combination of the localized
hard X-ray emission and the WD rotation (e.g., \cite{norton2007}).

The first method has been applied to four binary systems causing a classical nova to
date. Three of them (V2487 Oph; \cite{hernanz2002}, GK Per; \cite{hellier2004}, and CP
Pup; \cite{orio2009}) are considered to host a magnetic WD, while the remaining one
(V603 Aql; \cite{mukai2005}) is considered to host a non-magnetic WD that is
particularly bright as a nearby source at the distance of 360~pc \citep{hubble1927}. The
Fe\emissiontype{I} K$\alpha$ fluorescent line was detected from all these samples in the
post-nova phase, but the observations were made long after the outburst ($\sim$3~yr for
V2487 Oph, $\sim$60~yr for CP Pup, $\sim$80~yr for V603 Aql, and $\sim$100~yr for GK
Per). It is thus unclear how early the accretion process was reestablished. Observations
at a high cadence immediately after an outburst are obviously necessary. The second
method was applied to two classical novae to date, V603 Aql \citep{drechsel1983} and
V2491 Cyg \citep{page2010}. In the case of V2491 Cyg, the hard X-ray light curve started
to exhibit flickering no later than day 57 after the outburst, indicating that the
accretion process was reestablished at least this early. The third method has not been
unambiguously reported in the post-nova evolution.

\medskip

The purpose of this paper is to investigate the early reestablishement of the accretion
process in V2491 Cyg \citep{page2010} using the Fe\emissiontype{I} K$\alpha$ fluorescent
line method. We assemble archived data taken at a high cadence with the Swift satellite,
as well as our target-of-opportunity observations designed to obtain high
signal-to-noise ratio spectra in the Fe~K band (6.0--7.0~keV) at several post-nova
epochs with the Suzaku and XMM-Newton satellites. The object was observed
serendipitously in the pre-nova phase in X-rays \citep{ibarra2009}, and we also aim to
compare the X-ray spectra before and after the nova eruption.

The plan of this paper is as follows: In section~\ref{target}, we briefly summarize the
features of V2491 Cyg with ground-based and space-based observations. In
section~\ref{observation}, we describe the X-ray observations and the data processing
with the Suzaku, XMM-Newton, and Swift satellites. In section~\ref{analysis}, we
construct the X-ray spectra in the broad band (0.3--10~keV) as well as in the hard band
(4.0--10~keV) around the Fe~K complex at different time intervals from pre-nova to
post-nova phases. We then fit the spectra with simple models to characterize line and
continuum emission in the hard band. In section~\ref{discussion}, we develop discussion
to achieve the purpose mentioned above. Finally, the main results of this paper are
summarized in section~\ref{summary}.

\section{V2491 Cygni}\label{target}
\subsection{Ground-based observations}\label{target_ground}

The classical nova V2491 Cyg \citep{nakano2008,samus2008} was discovered on 2008 April
10.728~UT (54566.73~d in modified Julian date; MJD) in the constellation Cygnus at (RA,
Dec) $=$ ($\timeform{19h43m01.96s}$, $\timeform{+32D19'13.8''}$) in the equinox J2000.0.
We define the epoch of the discovery as the origin of time throughout this paper. V2491
Cyg is an extremely fast nova \citep{tomov2008a}, declining at a rate of $t_{2}$ $\sim$
4.6~d \citep{tomov2008b} and $t_{3}$ $\sim$ 16.8~d, where $t_{2}$ and $t_{3}$ are the
durations to fade respectively by 2 and 3~mag from the optical maximum. The distance to
V2491 Cyg was estimated to be $\sim$10.5~kpc \citep{helton2008} using an empirical
relation between the maximum magnitude and the rate of decline among classical novae
\citep{della1995}.

\subsection{Space-based observations}\label{target_space}

The discovery of V2491 Cyg triggered an intense monitoring campaign with Swift, which
continued for more than half a year \citep{kuulkers2008,osborne2008,page2008,page2010}.
Subsequent follow-up observations were also made by Suzaku and XMM-Newton. We present
results from these data sets in this paper. Below, we briefly summarize other results
published to date.

On day 9, Suzaku obtained an extremely flat X-ray spectrum extending up to 70~keV with a
6.7~keV emission line from Fe\emissiontype{XXV} \citep{takei2009}. The continuum
emission was speculated to be of non-thermal origin. On days 40 and 50, XMM-Newton
observations to obtain high-resolution grating X-ray spectra of the super-soft emission
were made (\cite{ness2008a,ness2008b,ness2010a,ness2011}). In these spectra,
blackbody-like continuum emission was found, together with broad absorption lines from
N\emissiontype{VI}, N\emissiontype{VII}, O\emissiontype{VII}, and O\emissiontype{VIII}
as well as emission lines from Ne\emissiontype{IX}, Ne\emissiontype{X}, and
Mg\emissiontype{XII}. \citet{ness2011} found that the majority of photospheric
absorption lines were blue-shifted by $\sim$3000~km~s$^{-1}$. Evidence for a possible
X-ray periodicity of $\sim$37~min was also present in the XMM-Newton light curve. Due to
the low number of observed cycles, however, it is not yet certain whether this was
caused by the WD rotation or red noise. If the periodicity was from rotation, it
strongly suggests that the outburst originated in an IP system.

Before the outburst, Swift observed an area coincidently including V2491 Cyg between
days $-$322 and $-$100. An X-ray source was significantly detected in five observations
(Swift J194302.1$+$321913; \cite{ibarra2008a,ibarra2008b,ibarra2009,page2010}), among
which the spectral shape was observed to change dramatically \citep{ibarra2009}. At the
position of V2491 Cyg, X-ray emission was also detected in the ROSAT all-sky survey
faint source catalogue (1RXS J194259.9$+$321940; \cite{voges2000}), the second ROSAT
PSPC catalogue\footnote{http://www.xray.mpe.mpg.de/rosat/rra/rospspc/}
(2RXP J194302.0$+$321912), and the XMM-Newton slew survey catalogue (XMMSL1
J194301.9$+$321911; \cite{saxton2008}). This makes V2491 Cyg one of the few examples of
classical novae with an X-ray detection prior to outburst. The pre-nova X-ray activity
suggests that it hosts a magnetic WD \citep{ibarra2009}.

\section{Observations and reduction}\label{observation}
\subsection{Suzaku}\label{observation_suzaku}

\begin{table*}[bt]
 \begin{center}
  \caption{List of our target-of-opportunity observations.}\label{table:observation}
  \begin{tabular*}{175mm}{@{\extracolsep{\fill}}clccccccc}
   \hline
   \phantom{.} & Mission/Detector & Obs ID      & Start time       & End time         & $\textit{t}$\footnotemark[$*$] & $\Delta{}\textit{t}$\footnotemark[$\dagger$] & Count rate\footnotemark[$\ddagger$] & \phantom{.} \\
   &                              &             & (UT)             & (UT)             & (d)             & (ks) & (s$^{-1}$) & \\
   \hline
   & Suzaku/XIS                   & 903001010   & 2008-04-19 15:21 & 2008-04-20 02:00 & \phantom{0}9.13 & 21   & 0.02  & \\
   &                              & 903001020   & 2008-05-09 08:39 & 2008-05-09 21:30 & 28.90           & 25   & 0.20  & \\
   \hline
   & XMM-Newton/EPIC-PN           & 0552270501  & 2008-05-20 14:24 & 2008-05-21 00:58 & 40.09           & 29   & 212.0 & \\
   &                              & 0552270601  & 2008-05-30 08:41 & 2008-05-30 17:05 & 49.81           & 27   & 38.4  & \\
   \hline
   \multicolumn{3}{@{}l@{}}{\hbox to 0pt{\parbox{175mm}{\footnotesize
   \par\noindent
   \footnotemark[$*$] Elapsed days in the middle of the observation from the discovery
   of V2491 Cyg (54566.73~d in MJD; \cite{nakano2008}).
   \par\noindent
   \footnotemark[$\dagger$] Net exposure time averaged over the operating CCDs.
   \par\noindent
   \footnotemark[$\ddagger$] The background-subtracted count rate in the 0.3--10~keV
   energy band over the operating CCDs.
   }\hss}}
  \end{tabular*}
 \end{center}
\end{table*}

\begin{figure}[tb]
 \begin{center}
  \FigureFile(85mm,85mm){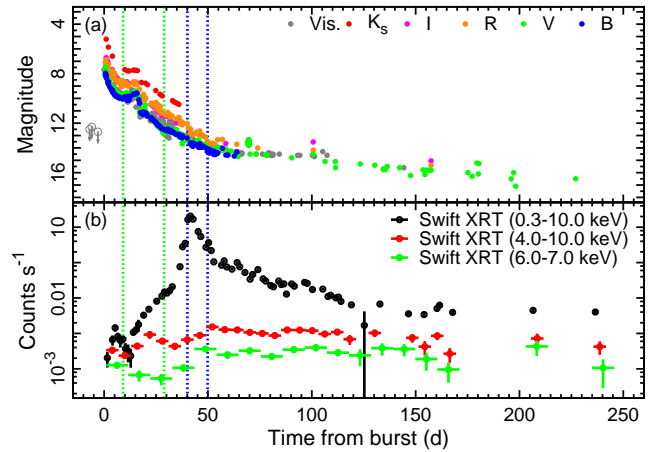}
 \end{center}
 \vspace{-3mm}
 \caption{Development of the (a) optical and (b) X-ray brightness of V2491 Cyg. The
 origin of the abscissa is 54566.73~d (MJD) when the classical nova was spotted
 \citep{nakano2008}. The time of the Suzaku and XMM-Newton observations are indicated by
 dotted lines with green and blue colors, respectively. (a) The \textit{B}-,
 \textit{V}-, \textit{R}-, \textit{I}-, \textit{K\rm{s}}-band, and visual magnitudes are
 shown with different colors. The upper limit for visual magnitudes before the outburst
 \citep{nakano2008} is indicated by open circles and arrows. (b) The
 background-subtracted X-ray count rates from Swift in the broad band (0.3--10~keV), the
 hard band (4.0--10~keV), and the Fe~K band (6.0--7.0~keV) with different colors. The
 optical photometry data are taken from the American Association of Variable Star
 Observers (AAVSO), the Variable Star Observers League in Japan (VSOLJ), and other
 ground-based observations
 \citep{nakano2008,lynch2008a,ashok2008,tomov2008a,tomov2008b,rudy2008,helton2008}. The
 infra-red photometry data were obtained by the Kanata TRISPEC team\footnotemark{}.
 }\label{figure:lcurve_optical}
\end{figure}
\footnotetext{http://kanatatmp.g.hatena.ne.jp/}

Following the X-ray detection by Swift during monitoring of the nova, we made two
target-of-opportunity observations using the Suzaku satellite on 2008 April 19 and May 9
(days 9 and 29; table~\ref{table:observation} and figure~\ref{figure:lcurve_optical}).
Suzaku has two instruments in operation \citep{mitsuda2007}: the X-ray Imaging
Spectrometer (XIS; \cite{koyama2007}) and the Hard X-ray Detector (HXD;
\cite{takahashi2007,kokubun2007}). The observations were aimed so as to put V2491 Cyg at
the XIS field center with different roll angles. Part of the data from the first
observation, emphasizing the HXD spectrum, was reported in \citet{takei2009}. In this
paper, we concentrate on the XIS data from both observations, focusing on the Fe~K band.

The XIS is equipped with four X-ray CCDs at the foci of four X-ray telescope modules
\citep{serlemitsos2007}. Three of them (XIS0, 2, and 3) are front-illuminated (FI) CCDs
sensitive in a 0.4--12~keV energy range. The remaining one (XIS1) is a back-illuminated
(BI) CCD sensitive at 0.2--12~keV. As of the observation dates, the absolute energy
scale is accurate to $\lesssim$10~eV, the energy resolution is 160--190~eV (FWHM) at
5.9~keV, and the total effective area is $\sim$400~cm$^{2}$ at 8~keV. Each XIS has a
format of 1024$\times$1024~pixels and covers a 18$\arcmin\times$18$\arcmin$ field of
view with an energy-independent half-power diameter of 1$\farcm$8--2$\farcm$3. XIS2 has
not been functional since 2006 November, thus we used the remaining three CCDs in our
observations. The XIS was operated in the normal clocking mode with a frame time of 8~s.
The spaced-row charge injection technique \citep{nakajima2008} were employed for the XIS
to rejuvenate its spectral resolution by filling the charge traps with artificially
injected electrons through CCD readouts.

The data were processed with the pipeline version 2.2.7.18, in which events were removed
during the South Atlantic anomaly passages, the night earth elevation angles below
5$^\circ$, and the day earth elevation angles below 20$^\circ$. For data reduction, we
used the High Energy Astrophysics Software (HEASoft) package\footnote{
http://heasarc.gsfc.nasa.gov/docs/software/lheasoft/} version 6.6.2 and the
calibration database version hxd20080201\_xis20080201\_xrt20070622\_xrs20060410. As a
result, the net exposure times of $\sim$21 and $\sim$25~ks were obtained for days 9 and
29, respectively.

\subsection{XMM-Newton}\label{observation_newton}

Following the flux increase in the soft X-ray band about a month after the nova
outburst, two target-of-opportunity observations were carried out using the XMM-Newton
satellite on 2008 May 20 and 30 (days 40 and 50; table~\ref{table:observation} and
figure~\ref{figure:lcurve_optical}). XMM-Newton has three instruments in operation
\citep{jansen2001}: the European Photon Imaging Camera (EPIC;
\cite{struder2001,turner2001}), the Reflection Grating Spectrometer (RGS;
\cite{herder2001}), and the Optical Monitor (OM; \cite{mason2001,telavera2009}). The
results obtained from the RGS data can be found in \citet{ness2008a}, \citet{ness2008b},
\citet{ness2010a}, and \citet{ness2011}. In this paper, we concentrate on the EPIC
data, focusing on the Fe~K band.

The EPIC is equipped with three X-ray CCD arrays at the foci of three X-ray telescopes.
Two of them use MOS-type CCDs (EPIC-MOS; \cite{turner2001}) sensitive in the
0.15--12~keV energy range, and the remaining one employs a PN-type CCD (EPIC-PN;
\cite{struder2001}) sensitive at 0.15--15~keV. The EPIC-MOS was operated in the small
window mode, while the EPIC-PN was in the timing mode. We do not use the EPIC-MOS
spectra of V2491 Cyg here for the following reasons: (1) they suffered severe pile-up,
and (2) a lower signal-to-noise dataset resulted because of the smaller effective area
than that of the EPIC-PN by a factor of $\sim$4 at the Fe~K band. Annulus source
extraction does not work to mitigate the pile-up issue in this case; the pile-up is
caused by intense soft X-ray band emission and it is difficult to obtain a sufficiently
strong signal for the fainter, harder emission in annular regions.

In the timing mode of the EPIC-PN, only a part of the CCD number 4 in the array is read
out in continuous clocking to mitigate pile-up. As a consequence, a one dimensional
(64$\times$1 pixels) image is obtained. The half-power diameter of the X-ray telescope
is $\sim$15$\arcsec$, which corresponds to 4~pixels. The absolute energy scale is
accurate to $\sim$50~eV at Fe~K band, the energy resolution is $\sim$150~eV (FWHM) at
6.4~keV \citep{guainazzi2010}, and the effective area is $\sim$600~cm$^{2}$ at 8~keV.

We used the Science Analysis Software (SAS)\footnote{http://xmm.esac.esa.int/sas/}
version 8.0.0 in data reduction. In the background light curve constructed
using the 10--12~keV energy band events, several spikes were recognized as flares. We
removed events taken during these flare intervals. The intervals were defined as time
bins with background count rates of more than 0.36 and 0.95~s$^{-1}$, respectively, for
the observations on days 40 and 50. These criteria were derived using a recursive
clipping algorithm (e.g., \cite{maughan2004}) to remove time bins having more background
counts than the average by 3-$\sigma$. As a result, net exposure times of $\sim$29 and
$\sim$27~ks were obtained for days 40 and 50, respectively.

\subsection{Swift}\label{observation_swift}

We retrieved the archived data taken by the Swift satellite both before and after the
outburst. Swift has three instruments in operation \citep{gehrels2004}: the Burst Alert
Telescope (BAT; \cite{barthelmy2005}), the X-Ray Telescope (XRT; \cite{burrows2005}) and
the Ultraviolet and Optical Telescope (UVOT; \cite{roming2005}). The XRT and UVOT
results were presented in \citet{page2010} and \citet{ibarra2009} for the post-nova and
pre-nova data, respectively. We reanalyze the XRT data focusing on the Fe~K band in this
paper.

The XRT is equipped with an X-ray CCD at the focus of an X-ray mirror module. The chip
has a format of 600$\times$600 pixels covering a 23$\farcm$6$\times$23$\farcm$6 field of
view at a telescope half-power diameter of $\sim$18$\arcsec$. The device is sensitive in
the range 0.3--10~keV, with an energy resolution of $\sim$180~eV (FWHM) at 5.9~keV as of
the observation dates. The effective area is $\sim$20~cm$^{2}$ at 8~keV
\citep{mukerjee2004}. The XRT was operated in the state that selects the clocking mode
automatically depending on the source count rate. The present data were mainly taken
with the Photon Counting (PC) and the Windowed Timing (WT) modes.  The PC mode provides
full imaging and spectroscopic resolution with a frame time of 2.5~s, while the WT mode
provides a one dimensional stacked image using the central 200$\times$10~pixels in
continuous clocking.

For data reduction, we used the HEASoft package\footnotemark[3]{} version 6.9 and the
calibration database version b20070924\_u20071106\_x20071101\_m20071023. The data were
processed using the \texttt{xrtpipeline} tool to produce level 2 event files with
standard screening criteria. As a result, total net exposure times of $\sim$18~ks and
$\sim$185~ks were obtained in the pre-nova and post-nova phases.

\section{Analysis}\label{analysis}
\subsection{Constructing spectra and light curves}\label{analysis_selection}
\subsubsection{Suzaku}\label{analysis_selection_suzaku}

\begin{figure*}[tb]
 \begin{center}
  \FigureFile(170mm,170mm){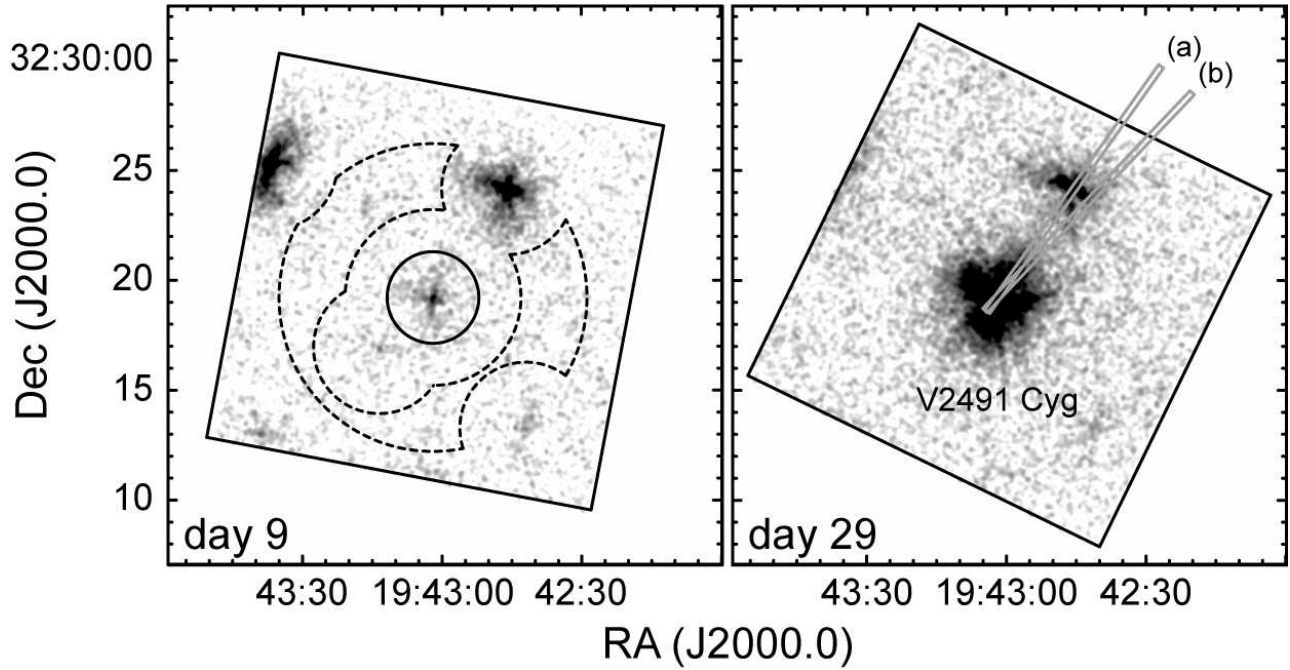}
 \end{center}
 \vspace{-3mm}
 \caption{Smoothed Suzaku XIS images of V2491 Cyg on days 9 and 29. Events recorded by
 the three CCDs in the 0.2--12~keV range were used, excluding those in the 5.0--7.0~keV
 range to eliminate the signals from the calibration sources at the corners. In the left
 panel, the XIS source extraction region is shown as a solid circle, while the background
 extraction region is shown using dashed lines. In the right panel, the source
 extraction regions for the EPIC-PN data are shown for days (a) 40 and (b) 50 as gray
 rectangles.
 }\label{figure:image_xis}
\end{figure*}

Figure~\ref{figure:image_xis} shows the smoothed XIS image in the 0.2--12~keV energy
band obtained on days 9 and 29. The astrometry of the XIS images were registered by
matching the position of V2491 Cyg with that by the optical observation
\citep{nakano2008}. The source events were accumulated from a circle with a radius of
120~pixels (2\farcm1) to maximize the signal-to-noise ratio (solid circle in
figure~\ref{figure:image_xis}), while the background events were from an annulus region
with inner and outer radii of 4$\arcmin$ and 7$\arcmin$, respectively. We masked
3$\arcmin$ radius circles around other detected sources in the background region.

\begin{figure}[tb]
 \begin{center}
  \FigureFile(85mm,85mm){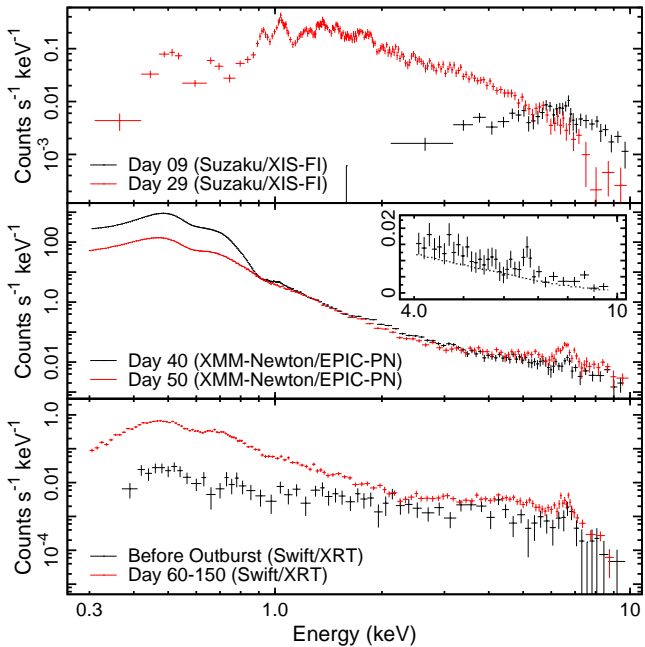}
 \end{center}
 \vspace{-3mm}
 \caption{Background-subtracted broad-band (0.3--10~keV) X-ray spectra of V2491 Cyg at
 six different epochs. The top, middle, and bottom panels show the spectra with
 Suzaku/XIS-FI, XMM-Newton/EPIC-PN, and Swift/XRT, respectively. The contamination by
 the nearby source in the XMM-Newton/EPIC-PN spectra on day 40 is shown as the dotted
 line in the inset of the middle panel.
 }\label{figure:spectrum_all}
\end{figure}

\begin{figure}[tb]
 \begin{center}
  \FigureFile(85mm,85mm){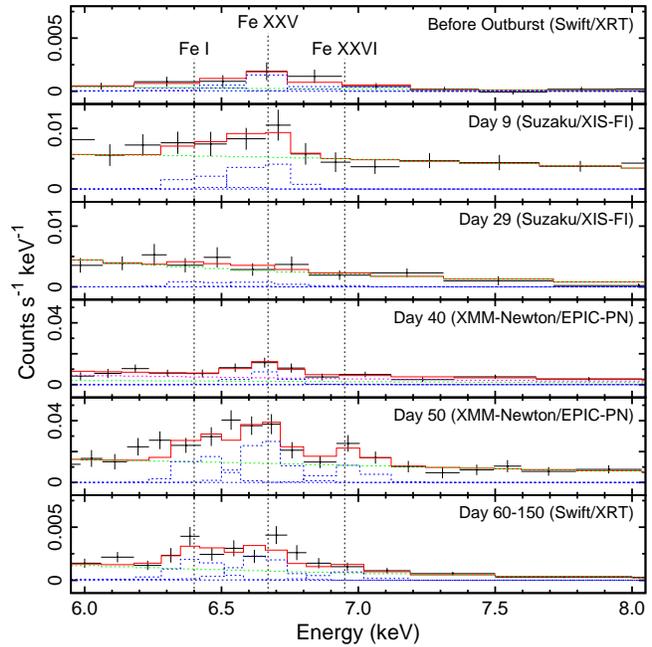}
 \end{center}
 \vspace{-3mm}
 \caption{Background-subtracted narrow-band (6.0--8.0~keV) X-ray spectra of V2491 Cyg,
 including the Fe~K complex, at six different epochs. The center energy of the
 three emission lines are indicated by dashed lines. Black pluses indicate the data,
 while the red solid histograms indicate the best-fit model. The model is decomposed
 into the power-law continuum (green dotted) and the three Gaussian line (blue dotted)
 components. The contamination by the nearby source in the XMM-Newton/EPIC-PN spectra on
 day 40 is shown as the dotted line in magenta.
 }\label{figure:spectrum_fe}
\end{figure}

The background-subtracted XIS spectra were constructed separately for days 9 and 29.
Figure~\ref{figure:spectrum_all} shows the broad-band (0.3--10~keV) spectra, while
figure~\ref{figure:spectrum_fe} shows the narrow-band (6.0--8.0~keV) spectra around the
Fe~K complex. The detector and mirror responses were generated using the
\texttt{xisrmfgen} and \texttt{xissimarfgen} tools \citep{ishisaki2007}, respectively.
The two XIS-FI spectra with nearly identical responses were merged, while the XIS-BI
spectrum was treated separately. The difference in the effective area due to different
off-axis angles were compensated between source and background regions using a method
described in \citet{hyodo2008}. The fraction of photon pile-up is negligible
($\lesssim$0.1\%) in both observations.

\subsubsection{XMM-Newton}\label{analysis_selection_newton}

In the one dimensional image of the EPIC-PN data obtained in the timing mode, V2491 Cyg
is centered near columns 36--37. The source events were accumulated from columns 35--38,
while the background events were from columns 3--4 and 61--62. We show
background-subtracted spectra both in the broad band (0.3--10~keV) and the narrow band
(6.0--8.0~keV) separately for days 40 and 50 (figures~\ref{figure:spectrum_all} and
\ref{figure:spectrum_fe}, respectively). The detector and mirror responses were
generated using standard SAS tools.

The V2491 Cyg spectrum on day 40 was unfortunately contaminated by a nearby source in
the source region (figure~\ref{figure:image_xis}). In the timing mode data, the image is
stacked in the readout direction, so their overlapping signals cannot be
separated. Therefore, we extracted the EPIC-MOS spectrum from the contaminating source
and evaluated the level of contamination. The hard-band (4.0--10~keV) spectrum was
fitted by a power-law model, and a statistically-acceptable fit was obtained for a
photon index of 1.5$\pm$0.7 and an X-ray flux of (7.6$\pm$1.5)$\times$10$^{-13}$
erg~cm$^{-2}$~s$^{-1}$ (90\% statistical uncertainty). We show the contribution of the
nearby source for the V2491 Cyg spectrum on day 40 in figure~\ref{figure:spectrum_all}
and \ref{figure:spectrum_fe}, in which we normalized the spectrum of the contaminating
source by taking into account the reduced effective area at an off-axis position. The
X-ray flux of the contamination source was estimated to be comparable with that of
V2491 Cyg in the hard band (4.0--10~keV). We checked all data sets from Suzaku,
XMM-Newton, and Swift observations, and no emission line was detected in the Fe~K
band. Therefore, the emission line in the V2491 Cyg spectrum on day 40 is entirely
attributable to V2491 Cyg itself. Hereafter, we remove the contamination by including
the best-fit model of the contaminating source in the spectral fitting of V2491 Cyg on
day 40. We note here that the signals of V2491 Cyg and the contamination source on day
50 can be clearly separated in the one dimensional image with different roll angles.

\subsubsection{Swift}\label{analysis_selection_swift}

For the XRT data, the source events were accumulated from a circle with a radius of 20
pixels (47$\arcsec$), while the background events were from an annulus with inner and
outer radii of 50 and 70~pixels (118$\arcsec$ and 165$\arcsec$), respectively. For
several data sets largely affected by photon pile-up, we masked the core in the source
extraction circle. The core radii were adaptively chosen to extract the source event
with a pile-up fraction of less than 1\% for each snapshot observation. The light curves
were constructed in the broad band (0.3--10~keV), the hard band (4.0--10~keV), and the
Fe~K band (6.0--7.0~keV), which are shown in figure~\ref{figure:lcurve_optical}. We
grouped several snapshot observations for each epoch. The telescope vignetting, the
reduced aperture due to pile-up masking and bad pixels were taken into account for the
aperture photometry.

In order to investigate the spectral development with sufficient photon statistics in
the Fe~K band, we stacked the pre-nova data between days $-$322 and $-$100 into one
spectrum and the post-nova data between days 60 and 150 into another, upon confirmation
that there is no significant change in the 6.0--7.0~keV light curve. The PC mode data
were used for this purpose to ensure sufficient energy resolution to resolve the three
Fe emission lines. We made the broad-band and the narrow-band spectra, which are
illustrated in figures~\ref{figure:spectrum_all} and \ref{figure:spectrum_fe}
respectively for the pre-nova and the post-nova phases.

\subsection{Spectral Modeling}\label{analysis_spectrum_iron}
\subsubsection{Phenomenological fitting}

\begin{table*}[tb]
 \begin{center}
  \caption{Best-fit parameter values from the phenomenological line fitting\footnotemark[$*$].}\label{table:fe_lines}
  \begin{tabular*}{\textwidth}{@{\extracolsep{\fill}}ccccccc}
   \hline\hline
   $\textit{t}$\footnotemark[$\dagger$] & \multicolumn{3}{c}{------ Flux (10$^{-6}$ photons cm$^{-2}$ s$^{-1}$) ------} & \multicolumn{3}{c}{------ Equivalent width (10$^{2}$~eV) ------} \\
   (d)                                           & Fe\emissiontype{I} & Fe\emissiontype{XXV} & Fe\emissiontype{XXVI} & Fe\emissiontype{I} & Fe\emissiontype{XXV} & Fe\emissiontype{XXVI} \\
   \hline
   \phantom{\footnotemark[$\ddagger$]}\{$-$322, $-$100\}\footnotemark[$\ddagger$] & $<$ 9.5          & 10.4 (2.1--18.5) & $<$ 13.7       & $<$ 7.9         & 5.0 (0.6--17.7)      & $<$ 13.3                 \\
   \phantom{0}9.13                                                                & $<$ 2.6          & 2.6 (1.1--4.2)   & $<$ 1.0        & $<$ 1.7         & 1.7 (0.5--2.9)       & $<$ 1.7                  \\
   28.90                                                                          & $<$ 1.5          & $<$ 1.5          & $<$ 1.1        & $<$ 2.3         & $<$ 2.3              & $<$ 2.9                  \\
   \phantom{\footnotemark[$\S$]}40.09\footnotemark[$\S$]                          & $<$ 1.3          & 3.0 (1.4--4.6)   & $<$ 1.6        & $<$ 3.5         & 6.9 (2.0--11.4)      & $<$ 4.0                  \\
   49.81                                                                          & 6.3 (3.6--9.4)   & 10.1 (6.9--13.0) & 4.3 (1.7--7.0) & 1.6 (0.9--2.4)  & 2.6 (1.8--3.7)       & 1.2 (0.4--2.2)           \\
   \phantom{\footnotemark[$\ddagger$]}\{60, 150\}\footnotemark[$\ddagger$]        & 10.2 (5.8--14.3) & 13.3 (8.4--19.1) & 5.4 (1.7--9.2) & 2.3 (1.2--3.7)  & 3.0 (1.8--4.9)       & 1.4 (0.4--2.9)           \\
   \hline
   \multicolumn{7}{@{}l@{}}{\hbox to 0pt{\parbox{180mm}{\footnotesize
   \par\noindent
   \footnotemark[$*$] The statistical uncertainties and upper limits indicate the
   90\% confidence ranges.
   \par\noindent
   \footnotemark[$\dagger$] Elapsed days from the discovery of V2491 Cyg (54566.73~d in
   MJD; \cite{nakano2008}).
   \par\noindent
   \footnotemark[$\ddagger$] The time \{X, Y\} indicates the time interval between days
   X and Y.
   \par\noindent
   \footnotemark[$\S$] Increased uncertainty due to the contaminating source is not included.
   }\hss}}
  \end{tabular*}
 \end{center}
\end{table*}

We first characterize line emission in the hard-band (4.0--10~keV) spectra. For
identifying the three lines in the Fe~K complex (figure~\ref{figure:spectrum_fe}), we
fitted the hard-band (4.0--10~keV) spectra using a phenomenological model composed of a
power-law continuum and three Gaussian line components. The Gaussian energy of each line
was tied among the six different spectra such that the line energies were not allowed to
vary between them. The Gaussian width was fixed to zero for all lines for all spectra:
as the expected width is negligible compared with the instrumental resolution. Other
parameters were allowed to vary independently for each spectrum.

\begin{figure}[tb]
 \begin{center}
  \FigureFile(85mm,85mm){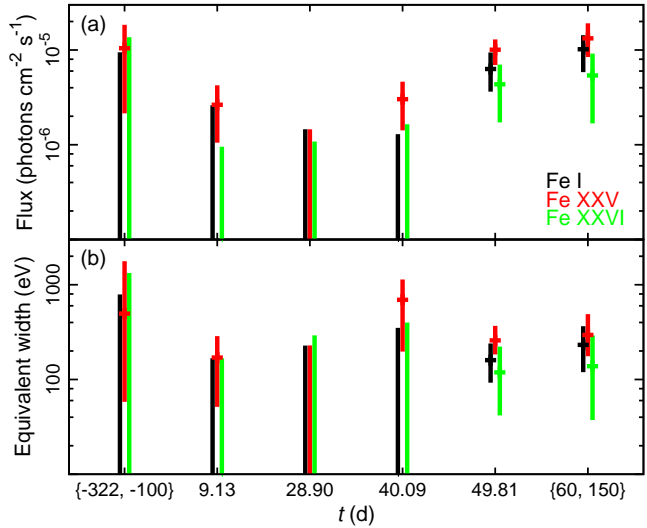}
 \end{center}
 \vspace{-3mm}
 \caption{Development of the (a) flux and (b) equivalent width of Fe\emissiontype{I}
 (black), Fe\emissiontype{XXV} (red), and Fe\emissiontype{XXVI} emission lines. The
 best-fit values and 90\% confidence ranges are shown at six different epochs. Increased
 uncertainty due to the contaminating source on day 40 is not included.
 }\label{figure:line_evolution}
\end{figure}

We obtained statistically acceptable results for the fitting, with the null hypothesis
probability for the $\chi^{2}$ value of $>$0.05. The best-fit energy centers for the
three Gaussian components are 6.42$\pm$0.08, 6.66$\pm$0.04, and 6.96$\pm$0.08~keV (90\%
statistical uncertainty), which are consistent respectively with the energies of the
K$\alpha$ fluorescent line from Fe\emissiontype{I}, the 1s$^{1}$S$_{0}$--2p$^{1}$P$_{1}$
resonance line from Fe\emissiontype{XXV}, and the Ly$\alpha$ line from
Fe\emissiontype{XXVI}. The flux and the equivalent width of each emission line are
tabulated in table~\ref{table:fe_lines}, in which 90\% upper limits are given for
undetected lines. The development of the Fe~K line intensities is also shown in
figure~\ref{figure:line_evolution}. The flux and the equivalent width on day 40 are
subject to further uncertainty due to the propagated uncertainty from the fitting of the
contaminating source, which is estimated as 2$\times$10$^{-7}$~photons~cm$^{-2}$~s$^{-1}$
and 300~eV, respectively.

\subsubsection{Plasma model fitting}\label{analysis_spectrum_suzaku}

\begin{table*}[tb]
 \begin{center}
  \caption{Best-fit parameter values from plasma model fitting\footnotemark[$*$].}\label{table:fe_band}
  \begin{tabular*}{\textwidth}{@{\extracolsep{\fill}}cccccc}
   \hline\hline
   $\textit{t}$\footnotemark[$\dagger$]                               & $\kt$           & $Z_{\rm{Fe}}$   & $\fx$\footnotemark[$\ddagger$]       & $\lx$\footnotemark[$\ddagger\S$] & $\chi^{2}_{\rm{red}} (\rm{d.o.f.})$ \\
   (d)                                                                & (keV)           & (solar)         & (10$^{-13}$~erg~cm$^{-2}$~s$^{-1}$)  & (10$^{33}$~erg~s$^{-1}$)         & \\
   \hline
   \phantom{\footnotemark[$\|$]}\{$-$322, $-$100\}\footnotemark[$\|$] & 7.6 (3.0--18.2) & 2.6 (0.8--7.0)  & 7.5 (2.1--9.2)                       & 9.9 (2.8--12.2)                  & 0.54 (16) \\
   \phantom{0}9.13                                                    & \multicolumn{5}{c}{(Plasma fitting omitted estimate because of dominant continuum from non-thermal emission.)} \\
   28.90                                                              & 2.1 (1.7--2.6)  & $<$ 0.1         & 5.5 (4.0--5.8)                       & 7.3 (5.3--7.7)                   & 0.77 (45) \\
   \phantom{\footnotemark[$\#$]}40.09\footnotemark[$\#$]              & 6.2 (3.0--10.9) & 1.3 (0.5--2.7)  & 2.5 (0.4--3.2)                       & 3.3 (0.6--4.2)                   & 0.65 (32) \\
   49.81                                                              & 9.6 (7.9--12.2) & 1.4 (1.0--1.8)  & 14.3 (12.8--15.3)                    & 18.9 (17.0--20.3)                & 1.34 (37) \\
   \phantom{\footnotemark[$\|$]}\{60, 150\}\footnotemark[$\|$]        & 7.3 (5.5--9.5)  & 1.1 (0.8--1.6)  & 16.7 (14.5--17.9)                    & 22.2 (19.3--23.8)                & 0.65 (30) \\
   \hline
   \multicolumn{4}{@{}l@{}}{\hbox to 0pt{\parbox{180mm}{\footnotesize
   \par\noindent
   \footnotemark[$*$] The statistical uncertainties and upper limits indicate the 90\%
   confidence ranges.
   \par\noindent
   \footnotemark[$\dagger$] Elapsed days from the discovery of V2491 Cyg (54566.73~d in
   MJD; \cite{nakano2008}).
   \par\noindent
   \footnotemark[$\ddagger$] These values are derived in the 4.0--10~keV energy band.
   \par\noindent
   \footnotemark[$\S$] A distance of 10.5~kpc \citep{helton2008} is assumed for the luminosity.
   \par\noindent
   \footnotemark[$\|$] The time \{X, Y\} indicates the time interval between days X and Y.
   \par\noindent
   \footnotemark[$\#$] Increased uncertainty due to the contaminating source is not included.
   }\hss}}
  \end{tabular*}
 \end{center}
\end{table*}

We next characterize the continuum and line emission using a thermal plasma emission
model. We fitted the hard-band (4.0--10~keV) spectra using an optically-thin isothermal
plasma component (APEC; \cite{smith2001}). This plasma component accounts for the
continuum as well as the emission lines from highly ionized Fe and other metals. We
fixed the chemical abundance to be solar \citep{anders1989} for all the elements other
than Fe for all the six different spectra. A narrow Gaussian component accounting for
the 6.4~keV emission line due to the Fe fluorescence was also included. The combined
model was multiplied with an interstellar absorption model (TBabs; \cite{wilms2000}).
The X-ray spectrum on day 9 was excluded from the plasma model analysis due to severe
contamination by non-thermal emission of V2491 Cyg that renders diagnosis of the softer
thermal component difficult. For fitting results for the data on day 9 using both plasma
and non-thermal models, see \citet{takei2009}.

We obtained statistically acceptable results for all the fitting. The best-fit plasma
temperature ($\kt$), the Fe abundance relative to solar ($Z_{\rm{Fe}}$), the detected
X-ray flux ($\fx$), and the absorption-corrected luminosity ($\lx$) in the hard band
(4.0--10~keV) are summarized in table~\ref{table:fe_band}. A distance of 10.5~kpc was
assumed \citep{helton2008}. The absorption column density ($\nh$) was fixed to be
2.2$\times$10$^{21}$~cm$^{-2}$ based on the analysis of the quiescent phase data by
\citet{ibarra2009}. By extrapolating the best-fit model, we estimated the X-ray
luminosity in the 2.0--10~keV energy band to be in the order of
$\sim$10$^{34}$~erg~s$^{-1}$. The fitting results on day 40 are again subject to
increased uncertainty in the fitting uncertainty of the contaminating source. It is
estimated that $\pm$1.5~keV in $\kt$, $\pm$2.5~solar in $Z_{\rm{Fe}}$,
$\pm$1.0$\times$10$^{-13}$~erg~cm$^{-2}$~s$^{-1}$ in $\fx$, and
1.3$\times$10$^{33}$~erg~s$^{-1}$ in $\lx$.

\section{Discussion}\label{discussion}
\subsection{Phases of X-ray emission}\label{discussion_phase}

V2491 Cyg shows completely different light curves for the soft and hard parts of the
X-ray band (figure~\ref{figure:lcurve_optical}; see also figures~1 and 4 in
\cite{page2010}). The soft-band flux was negligible immediately after the outburst, but
continued to increase to reach a peak at day $\sim$40, then faded rapidly until day
$\sim$50 and more slowly thereafter. This behavior is commonly seen in the soft X-ray
light curves of other classical novae. The increase in soft-band brightness before the
peak is interpreted in terms of an increasing plasma temperature (e.g.,
\cite{osborne2011}) and/or a decreasing extinction column as the ejecta expands, while
the decrease in brightness after the peak is interpreted as the shrinkage of the
photosphere after the fuel, in the form of H-rich accreted material, is consumed.

In contrast to the soft band, the hard-band light curve is relatively stable
(figure~\ref{figure:lcurve_optical}; see also figures~1 and 4 in \cite{page2010}).
Despite the overall stability, however, we argue that the hard-band development can be
divided into two phases (earlier and later phases hereafter) separated at day $\sim$40
based on the following lines of evidence: (1) In the Fe~K band light curve
(figure~\ref{figure:lcurve_optical}), the flux changes gradually in the earlier phase
with a minimum on day $\sim$30, whereas the flux is stable in the later phase. (2) In
the narrow-band spectra (figure~\ref{figure:spectrum_fe}), the Fe\emissiontype{I}
K$\alpha$ fluorescent line at 6.4~keV is only detected with a significance greater than
the 3-$\sigma$ level in the later phase (figure~\ref{figure:line_evolution}). (3) In the
plasma model fitting results (table~\ref{table:fe_band}), the Fe abundance relative to H
is different between the two phases. In the earlier phase, a stringent upper limit to
the Fe abundance was obtained for the spectra on day 29 of $Z_{\rm{Fe}}$ $<$ 0.1, which
is reasonable from the lack of 6.7~keV lines despite the plasma temperature
($\sim$2.1~keV) being close to the maximum formation temperature of the line (e.g.,
\cite{rothenflug1985}). In the later phase, the two lines from highly ionized Fe are
clearly detected (figure~\ref{figure:spectrum_fe}) with a corresponding Fe abundance of
$Z_{\rm{Fe}}$ $\sim$ 1. Here, we note that the Fe abundance values are subject to change
with the unconstrained chemical composition of other elements, which contribute to
continuum emission in the X-ray band. However, we consider that the comparative values
between two different spectra are valid as long as the chemical abundance of the plasma
is assumed to be the same.

\subsection{Origins of the X-ray emission}\label{discussion_origin}

The two distinctive phases in the hard-band evolution of V2491 Cyg suggest that there
are two different production mechanisms operating in each phase. We consider that plasma
generated by shocks in the ejecta is responsible for the earlier phase, while emission
by accretion shocks in the rekindled accretion process is responsible for the later
phase.

Hard components from classical novae were first reported in V838 Her \citep{lloyd1992}.
Such hard emission has been established as a common phenomenon among classical novae by
recent intensive monitoring observations with Swift (e.g.,
\cite{bode2006,ness2009b,page2009,page2010,osborne2011}) and subsequent follow-up
observations with XMM-Newton, Suzaku, and others (e.g.,
\cite{sokoloski2006,nelson2008,tsujimoto2009}). Viable interpretations for the hard
components include the plasma produced by internal shocks; i.e., ejecta with different
velocities collide with each other \citep{friedjung1987,mukai2001}. The X-ray spectra
are represented by optically-thin thermal plasma emission with a plasma temperature of
several keV (e.g., \cite{lloyd1992,mukai2001,orio2001,tsujimoto2009}). For unknown
reasons, Fe is often depleted in the chemical composition (e.g, V382 Vel;
\cite{orio2001}, V458 Vul; \cite{tsujimoto2009}). These features are quite similar to
those found in the hard component in the earlier phase of V2491 Cyg.

On the other hand, the hard component in the later phase is more similar to that
commonly seen in IPs. The defining characteristics of IP X-ray emission include the
following (e.g., \cite{ezuka1999}): (1) The Fe emission lines are clearly seen at 6.4,
6.7, and 7.0~keV respectively from Fe\emissiontype{I}, Fe\emissiontype{XXV}, and
Fe\emissiontype{XXVI}. (2) The continuum emission is very flat, which is represented by
thermal bremsstrahlung of a temperature of $\sim$10~keV. (3) The X-ray luminosity in the
2.0--10~keV energy band is $\sim$10$^{33}$~erg~s$^{-1}$ (e.g.,
\cite{chanmugam1991,norton1989,warner2003}), which is much brighter than that of
non-magnetic or weakly magnetized WDs and polars (e.g.,
\cite{mukai1993,warner2003,baskill2005,byckling2010}). The hard-band luminosity of V2491
Cyg ($\sim$10$^{34}$~erg~s$^{-1}$) is even larger than the typical IPs. This may be due
to a higher accretion rate in the system. (4) The harder band ($>$2~keV) emission
dominates the spectrum in IPs. In contrast, the softer band ($<$2~keV) emission
dominates the spectrum in polars. (5) The modulation due to the WD spin and orbital
motions are shown in the X-ray light curve (see \cite{ness2011} for results of the
timing analysis for V2491 Cyg). In IPs, the high temperature plasma responsible for the
flat continuum and the highly ionized Fe lines is caused by a shock at the base of the
accretion funnel. The plasma illuminates the WD surface, which produces a prominent
Fe\emissiontype{I} K$\alpha$ fluorescent line. All these features are seen in the
hard-band emission in the later phase of V2491 Cyg.

It is important to note three things in the results. The first is the equivalent width
of the Fe\emissiontype{I} line found in the later phase. The observed equivalent width
($\gtrsim$100~eV) is comparable to those observed in IPs (e.g., \cite{ezuka1999}). It
can be explained by fluorescence at the WD surface, or at the surface of the accretion
disk, by X-ray emitting plasma at the base of the accretion funnel in the proximity of
the WD (e.g., \cite{ishida1991,george1991}). It corresponds to the ``infinite plane''
fluorescence included in the calculations by \citet{drake2008}. However, such a large
equivalent width cannot be explained for cases in which (1) the plasma heated by ejecta
shocks illuminates the WD surface, or (2) the plasma heated by ejecta or accretion
shocks illuminates cold material in the ejecta. In case (1), the subtended angle of the
WD surface is too small ($\Delta\Omega$/4$\pi$ $\sim$ 10$^{-12}$, assuming that the
X-ray emitting region has expanded for 50~d at a constant speed of 3000~km~s$^{-1}$). In
case (2), the cold material is too thin to produce the Fe fluorescence with a large
equivalent width. Although the cold matter subtends the entire sphere (4$\pi$
steradian), the ejecta with a hydrogen column density of $\sim$10$^{21}$~cm$^{-2}$
\citep{page2010} can yield the equivalent width of only $\lesssim$10~eV for $\sim$10~keV
thermal plasma X-rays (see equation 4 in \cite{tsujimoto2005}). This is in line with our
interpretation that the hard component in the later phase is from the accretion process
in the WD.

The second is that the hard-band features in the pre-nova phase of V2491 Cyg resemble
those in the later phase of the post-nova spectra more than those in the earlier phase
regarding the presence of Fe emission lines (figure~\ref{figure:spectrum_fe}) and the
spectral hardness (table~\ref{table:fe_band}). This suggests that the origin of dominant
X-ray emission is the same between the pre-nova and the later post-nova phases. A
notable difference can be found in the hard-band flux. It is significantly higher in
the late post-nova phase after day 50 than in the pre-nova phase. We speculate that the
accretion rate in the late post-nova phase may have been higher than that in the
pre-nova phase, which might be expected due to the heating of the secondary star by the
outburst and the heated WD \citep{schaefer2010b}.

The third is that V2487 Oph \citep{hernanz2002}, GK Per \citep{hellier2004}, and CP Pup
\citep{orio2009}, which are the other three among four classical novae with a
Fe\emissiontype{I} fluorescence detection in the post-nova phase, also share
characteristics with IPs such as high X-ray luminosity in the quiescent phase (e.g.,
\cite{cordova1981,balman1995,hernanz2002}) and the rapidity of the optical decline
($t_{3}$ is $\sim$8~d for V2487 Oph; \cite{liller1999,pagnotta2009}, $\sim$13~d for GK
Per; \cite{downes2000}, and $\sim$6.5~d for CP Pup; \cite{payne1957,orio2009}).

\subsection{Time-scale of accretion reestablishment}\label{discussion_accretion}

Using the Fe emission lines, we argued that the accretion process in V2491 Cyg was
reestablished by at least $\sim$50 days after the outburst. \citet{page2010} argued that
the accretion was rekindled no later than day 57 using the X-ray flux flickering method.
These independent methods lead to the same conclusion on the time scale for the
accretion process to resume after the nova.

The time for the reestablishement of the accretion process coincides with the time when
the soft-band emission starts to decline (figure~\ref{figure:lcurve_optical}). This is
consistent with the following scenario. The accretion disk was likely destroyed by the
blast \citep{drake2010}. Immediately after the outburst, the strong radiation pressure
inhibits rejuvenation of the accretion disk and likely exceeds the ram pressure of the
accretion flow once reestablished, thereby preventing accretion to occur. When the
H-rich fuel had been consumed on the WD surface by nuclear burning and the
radiation pressure started to drop, the accretion resumed. We further speculate that the
magnetic field on the WD is responsible for the short time scale of rekindled accretion.
When mass accretion from the companion star was switched on after a post-nova interval,
the viscous time scale of the accretion disk would have to elapse before gas could again
reach the WD surface (e.g., \cite{frank1992}). For a magnetized WD, accretion could
resume earlier as less disk spreading is required. The scenario will be tested by
examining the coincidence of the decline in the soft-band flux and the emergence of the
Fe\emissiontype{I} emission line by high-cadence X-ray observations of classical novae
in the future.

\section{Summary}\label{summary}

We conducted a spectroscopic study of the classical nova V2491 Cyg in the hard X-ray
band (4.0--10~keV) using our target-of-opportunity Suzaku and XMM-Newton data, as well
as archived Swift data. X-ray spectra were obtained at five post-nova epochs on days 9,
29, 40, 50, and 60--150, in addition to a pre-nova interval between days $-$322 and
$-$100. In the time series of the spectra, we found remarkable changes in the X-ray
emission: (a) In the pre-nova phase and on day 9, the 6.7~keV emission line from
Fe\emissiontype{XXV} was significantly detected. The pre-nova spectra represented
thermal plasma emission with an Fe abundance of 0.8--7.0~solar. (b) On day 29, no Fe
lines were detected in the spectra. The Fe abundance of the dominant emission was
constrained to be less than 0.1~solar. (c) On day 40, the 6.7~keV emission line was
detected again. (d) In a later phase, three emission lines at 6.4, 6.7, and 7.0~keV from
Fe\emissiontype{I}, Fe\emissiontype{XXV}, Fe\emissiontype{XXVI}, respectively, were
found in the spectra. The Fe abundance and the flux in the hard X-ray band were
significantly larger than those of the earlier phase. We also confirmed the long-term
behavior of X-ray emission in the Fe~K band (6.0--7.0~keV), in which the flux changes
gradually in the earlier phase with a minimum on day $\sim$30 and it is stable in the
later phase.

Based on these results, we conclude that (1) the post-nova evolution in the X-ray light
curve can be divided into two different phases separated at day $\sim$40, (2) it is most
likely that shocks within the nova ejecta are responsible for the X-ray emission in the
earlier phase, while shocks in the rekindled accretion are responsible for the later
phase X-rays, and (3) the accretion process is considered to be reestablished by at
least day $\sim$50 when the Fe\emissiontype{I} emission line emerged, which is a common
feature for accretion shock emission from IPs.

\bigskip

The authors appreciate the reviewer for useful suggestions. The authors also thank the
Suzaku and XMM-Newton telescope managers for allocating telescope times for our
target-of-opportunity programs. D.\,T. expresses gratitude for the hospitality at the
European Space Astronomy Centre (ESAC) during the course of this work. We thank the
XMM-Newton Science Operations Centre (SOC) for technical advice and especially Matteo
Guainazzi for his continuous support. The near infra-red data were provided by the
Kanata TRISPEC team prior to publication, while the optical data were taken from the
AAVSO International Database and the VSOLJ Observation Database contributed by observers
worldwide.

This research has made use of data obtained from Data ARchives and Transmission System
(DARTS), provided by the Center for Science-satellite Operation and Data Archives (C-SODA)
at the Institute of Space and Astronautical Science (ISAS) of Japan Aerospace Exploration
Agency (JAXA), the High Energy Astrophysics Science Archive Research Center (HEASARC),
provided by the National Aeronautics and Space Administration (NASA) Goddard Space
Flight Center (GSFC), and of the SIMBAD database, operated at CDS, Strasbourg, France.

D.\,T. is financially supported by the Japan Society for the Promotion of Science
(JSPS). J.\,J.\,D. was supported by the NASA contract NAS8-39073 to the Chandra X-ray
Observatory Center (CXC). J.\,P.\,O. acknowledges the support of the Science and
Technology Facilities Council (STFC).

\bibliographystyle{ms}
\bibliography{ms}

\begin{thebibliography}{98}
\expandafter\ifx\csname natexlab\endcsname\relax\def\natexlab#1{#1}\fi

\bibitem[{{Anders} \& {Grevesse}(1989)}]{anders1989}
{Anders}, E. \& {Grevesse}, N. 1989, \gca, 53, 197

\bibitem[{{Ashok} {et~al.}(2008){Ashok}, {Banerjee}, {Joshi}, {Naik}, {Forne},
  \& {Santangelo}}]{ashok2008}
{Ashok}, N.~M., {Banerjee}, D.~P.~K., {Joshi}, V., {et~al.} 2008, Central
  Bureau Electronic Telegrams, 1354, 1

\bibitem[{{Balman} {et~al.}(1995){Balman}, {Orio}, \& {Ogelman}}]{balman1995}
{Balman}, S., {Orio}, M., \& {Ogelman}, H. 1995, \apj, 449, L47

\bibitem[{{Barthelmy} {et~al.}(2005){Barthelmy}, {Barbier}, {Cummings},
  {Fenimore}, {Gehrels}, {Hullinger}, {Krimm}, {Markwardt}, {Palmer},
  {Parsons}, {Sato}, {Suzuki}, {Takahashi}, {Tashiro}, \&
  {Tueller}}]{barthelmy2005}
{Barthelmy}, S.~D., {Barbier}, L.~M., {Cummings}, J.~R., {et~al.} 2005, Space
  Science Reviews, 120, 143

\bibitem[{{Baskill} {et~al.}(2005){Baskill}, {Wheatley}, \&
  {Osborne}}]{baskill2005}
{Baskill}, D.~S., {Wheatley}, P.~J., \& {Osborne}, J.~P. 2005, \mnras, 357, 626

\bibitem[{{Bode} \& {Evans}(2008)}]{bode2008}
{Bode}, M.~F. \& {Evans}, A. 2008, {Classical Novae}, ed. {Bode, M.~F.~\&
  Evans, A.} ({Cambridge: Cambridge University Press})

\bibitem[{{Bode} {et~al.}(2006){Bode}, {O'Brien}, {Osborne}, {Page},
  {Senziani}, {Skinner}, {Starrfield}, {Ness}, {Drake}, {Schwarz}, {Beardmore},
  {Darnley}, {Eyres}, {Evans}, {Gehrels}, {Goad}, {Jean}, {Krautter}, \&
  {Novara}}]{bode2006}
{Bode}, M.~F., {O'Brien}, T.~J., {Osborne}, J.~P., {et~al.} 2006, \apj, 652,
  629

\bibitem[{{Burrows} {et~al.}(2005){Burrows}, {Hill}, {Nousek}, {Kennea},
  {Wells}, {Osborne}, {Abbey}, {Beardmore}, {Mukerjee}, {Short}, {Chincarini},
  {Campana}, {Citterio}, {Moretti}, {Pagani}, {Tagliaferri}, {Giommi},
  {Capalbi}, {Tamburelli}, {Angelini}, {Cusumano}, {Br{\"a}uninger}, {Burkert},
  \& {Hartner}}]{burrows2005}
{Burrows}, D.~N., {Hill}, J.~E., {Nousek}, J.~A., {et~al.} 2005, Space Science
  Reviews, 120, 165

\bibitem[{{Byckling} {et~al.}(2010){Byckling}, {Mukai}, {Thorstensen}, \&
  {Osborne}}]{byckling2010}
{Byckling}, K., {Mukai}, K., {Thorstensen}, J.~R., \& {Osborne}, J.~P. 2010,
  \mnras, 408, 2298

\bibitem[{{Chanmugam} {et~al.}(1991){Chanmugam}, {Ray}, \&
  {Singh}}]{chanmugam1991}
{Chanmugam}, G., {Ray}, A., \& {Singh}, K.~P. 1991, \apj, 375, 600

\bibitem[{{Cordova} {et~al.}(1981){Cordova}, {Jensen}, \&
  {Nugent}}]{cordova1981}
{Cordova}, F.~A., {Jensen}, K.~A., \& {Nugent}, J.~J. 1981, \mnras, 196, 1

\bibitem[{{Cropper}(1990)}]{cropper1990}
{Cropper}, M. 1990, \ssr, 54, 195

\bibitem[{{della Valle} \& {Livio}(1995)}]{della1995}
{della Valle}, M. \& {Livio}, M. 1995, \apj, 452, 704

\bibitem[{{den Herder} {et~al.}(2001){den Herder}, {Brinkman}, {Kahn},
  {Branduardi-Raymont}, {Thomsen}, {Aarts}, {Audard}, {Bixler}, {den Boggende},
  {Cottam}, {Decker}, {Dubbeldam}, {Erd}, {Goulooze}, {G{\"u}del}, {Guttridge},
  {Hailey}, {Janabi}, {Kaastra}, {de Korte}, {van Leeuwen}, {Mauche},
  {McCalden}, {Mewe}, {Naber}, {Paerels}, {Peterson}, {Rasmussen}, {Rees},
  {Sakelliou}, {Sako}, {Spodek}, {Stern}, {Tamura}, {Tandy}, {de Vries},
  {Welch}, \& {Zehnder}}]{herder2001}
{den Herder}, J.~W., {Brinkman}, A.~C., {Kahn}, S.~M., {et~al.} 2001, \aap,
  365, L7

\bibitem[{{Downes} \& {Duerbeck}(2000)}]{downes2000}
{Downes}, R.~A. \& {Duerbeck}, H.~W. 2000, \aj, 120, 2007

\bibitem[{{Drake} {et~al.}(2008){Drake}, {Ercolano}, \& {Swartz}}]{drake2008}
{Drake}, J.~J., {Ercolano}, B., \& {Swartz}, D.~A. 2008, \apj, 678, 385

\bibitem[{{Drake} \& {Orlando}(2010)}]{drake2010}
{Drake}, J.~J. \& {Orlando}, S. 2010, \apjl, 720, L195

\bibitem[{{Drechsel} {et~al.}(1983){Drechsel}, {Rahe}, {Wargau}, {Seward}, \&
  {Wang}}]{drechsel1983}
{Drechsel}, H., {Rahe}, J., {Wargau}, W., {Seward}, F.~D., \& {Wang}, Z.~R.
  1983, \aap, 126, 357

\bibitem[{{Ezuka} \& {Ishida}(1999)}]{ezuka1999}
{Ezuka}, H. \& {Ishida}, M. 1999, \apj, 120, S277

\bibitem[{{Frank} {et~al.}(1992){Frank}, {King}, \& {Raine}}]{frank1992}
{Frank}, J., {King}, A., \& {Raine}, D. 1992, {Accretion power in
  astrophysics.}, ed. {Frank, J., King, A., \& Raine, D.}

\bibitem[{{Friedjung}(1987)}]{friedjung1987}
{Friedjung}, M. 1987, \aap, 180, 155

\bibitem[{{Gehrels} {et~al.}(2004){Gehrels}, {Chincarini}, {Giommi}, {Mason},
  {Nousek}, {Wells}, {White}, {Barthelmy}, {Burrows}, {Cominsky}, {Hurley},
  {Marshall}, {M{\'e}sz{\'a}ros}, {Roming}, {Angelini}, {Barbier}, {Belloni},
  {Campana}, {Caraveo}, {Chester}, {Citterio}, {Cline}, {Cropper}, {Cummings},
  {Dean}, {Feigelson}, {Fenimore}, {Frail}, {Fruchter}, {Garmire}, {Gendreau},
  {Ghisellini}, {Greiner}, {Hill}, {Hunsberger}, {Krimm}, {Kulkarni}, {Kumar},
  {Lebrun}, {Lloyd-Ronning}, {Markwardt}, {Mattson}, {Mushotzky}, {Norris},
  {Osborne}, {Paczynski}, {Palmer}, {Park}, {Parsons}, {Paul}, {Rees},
  {Reynolds}, {Rhoads}, {Sasseen}, {Schaefer}, {Short}, {Smale}, {Smith},
  {Stella}, {Tagliaferri}, {Takahashi}, {Tashiro}, {Townsley}, {Tueller},
  {Turner}, {Vietri}, {Voges}, {Ward}, {Willingale}, {Zerbi}, \&
  {Zhang}}]{gehrels2004}
{Gehrels}, N., {Chincarini}, G., {Giommi}, P., {et~al.} 2004, \apj, 611, 1005

\bibitem[{{George} \& {Fabian}(1991)}]{george1991}
{George}, I.~M. \& {Fabian}, A.~C. 1991, \mnras, 249, 352

\bibitem[{{Guainazzi} {et~al.}(2010){Guainazzi}, {Kirsch}, F., \&
  {de~Juan~Ovelar}}]{guainazzi2010}
{Guainazzi}, M., {Kirsch}, M., F., H., \& {de~Juan~Ovelar}, M. 2010,
  {Evaluation of the spectral calibration accuracy in EPIC-pn fast modes},
  version 1.2

\bibitem[{{Hellier} \& {Mukai}(2004)}]{hellier2004}
{Hellier}, C. \& {Mukai}, K. 2004, \mnras, 352, 1037

\bibitem[{{Helton} {et~al.}(2008){Helton}, {Woodward}, {Vanlandingham}, \&
  {Schwarz}}]{helton2008}
{Helton}, L.~A., {Woodward}, C.~E., {Vanlandingham}, K., \& {Schwarz}, G.~J.
  2008, Central Bureau Electronic Telegrams, 1379, 1

\bibitem[{{Hernanz} \& {Sala}(2002)}]{hernanz2002}
{Hernanz}, M. \& {Sala}, G. 2002, Science, 298, 393

\bibitem[{{Hubble} \& {Duncan}(1927)}]{hubble1927}
{Hubble}, E.~P. \& {Duncan}, J.~C. 1927, \apj, 66, 59

\bibitem[{{Hyodo} {et~al.}(2008){Hyodo}, {Tsujimoto}, {Hamaguchi}, {Koyama},
  {Kitamoto}, {Maeda}, {Tsuboi}, \& {Ezoe}}]{hyodo2008}
{Hyodo}, Y., {Tsujimoto}, M., {Hamaguchi}, K., {et~al.} 2008, \pasj, 60, S85

\bibitem[{{Ibarra} \& {Kuulkers}(2008)}]{ibarra2008a}
{Ibarra}, A. \& {Kuulkers}, E. 2008, The Astronomer's Telegram, 1473, 1

\bibitem[{{Ibarra} {et~al.}(2008){Ibarra}, {Kuulkers}, {Beardmore}, {Evans},
  {Mukai}, {Ness}, {Orio}, {Osborne}, {Page}, {Saxton}, {Starrfield}, \&
  {Tueller}}]{ibarra2008b}
{Ibarra}, A., {Kuulkers}, E., {Beardmore}, A., {et~al.} 2008, The Astronomer's
  Telegram, 1478, 1

\bibitem[{{Ibarra} {et~al.}(2009){Ibarra}, {Kuulkers}, {Osborne}, {Page},
  {Ness}, {Saxton}, {Baumgartner}, {Beckmann}, {Bode}, {Hernanz}, {Mukai},
  {Orio}, {Sala}, {Starrfield}, \& {Wynn}}]{ibarra2009}
{Ibarra}, A., {Kuulkers}, E., {Osborne}, J.~P., {et~al.} 2009, \aap, 497, L5

\bibitem[{{Ishida} {et~al.}(1991){Ishida}, {Silber}, {Bradt}, {Remillard},
  {Makishima}, \& {Ohashi}}]{ishida1991}
{Ishida}, M., {Silber}, A., {Bradt}, H.~V., {et~al.} 1991, \apj, 367, 270

\bibitem[{{Ishisaki} {et~al.}(2007){Ishisaki}, {Maeda}, {Fujimoto}, {Ozaki},
  {Ebisawa}, {Takahashi}, {Ueda}, {Ogasaka}, {Ptak}, {Mukai}, {Hamaguchi},
  {Hirayama}, {Kotani}, {Kubo}, {Shibata}, {Ebara}, {Furuzawa}, {Iizuka},
  {Inoue}, {Mori}, {Okada}, {Yokoyama}, {Matsumoto}, {Nakajima}, {Yamaguchi},
  {Anabuki}, {Tawa}, {Nagai}, {Katsuda}, {Hayashida}, {Bamba}, {Miller},
  {Sato}, \& {Yamasaki}}]{ishisaki2007}
{Ishisaki}, Y., {Maeda}, Y., {Fujimoto}, R., {et~al.} 2007, \pasj, 59, S113

\bibitem[{{Jansen} {et~al.}(2001){Jansen}, {Lumb}, {Altieri}, {Clavel}, {Ehle},
  {Erd}, {Gabriel}, {Guainazzi}, {Gondoin}, {Much}, {Munoz}, {Santos},
  {Schartel}, {Texier}, \& {Vacanti}}]{jansen2001}
{Jansen}, F., {Lumb}, D., {Altieri}, B., {et~al.} 2001, \aap, 365, L1

\bibitem[{{Kokubun} {et~al.}(2007){Kokubun}, {Makishima}, {Takahashi},
  {Murakami}, {Tashiro}, {Fukazawa}, {Kamae}, {Madejski}, {Nakazawa},
  {Yamaoka}, {Terada}, {Yonetoku}, {Watanabe}, {Tamagawa}, {Mizuno}, {Kubota},
  {Isobe}, {Takahashi}, {Sato}, {Takahashi}, {Hong}, {Kawaharada}, {Kawano},
  {Mitani}, {Murashima}, {Suzuki}, {Abe}, {Miyawaki}, {Ohno}, {Tanaka},
  {Yanagida}, {Itoh}, {Ohnuki}, {Tamura}, {Endo}, {Hirakuri}, {Hiruta},
  {Kitaguchi}, {Kishishita}, {Sugita}, {Takahashi}, {Takeda}, {Enoto},
  {Hirasawa}, {Katsuta}, {Matsumura}, {Onda}, {Sato}, {Ushio}, {Ishikawa},
  {Murase}, {Odaka}, {Suzuki}, {Yaji}, {Yamada}, {Yamasaki}, {Yuasa}, \& {The
  Hxd Team}}]{kokubun2007}
{Kokubun}, M., {Makishima}, K., {Takahashi}, T., {et~al.} 2007, \pasj, 59, S53

\bibitem[{{Koyama} {et~al.}(2007){Koyama}, {Tsunemi}, {Dotani}, {Bautz},
  {Hayashida}, {Tsuru}, {Matsumoto}, {Ogawara}, {Ricker}, {Doty}, {Kissel},
  {Foster}, {Nakajima}, {Yamaguchi}, {Mori}, {Sakano}, {Hamaguchi},
  {Nishiuchi}, {Miyata}, {Torii}, {Namiki}, {Katsuda}, {Matsuura}, {Miyauchi},
  {Anabuki}, {Tawa}, {Ozaki}, {Murakami}, {Maeda}, {Ichikawa}, {Prigozhin},
  {Boughan}, {Lamarr}, {Miller}, {Burke}, {Gregory}, {Pillsbury}, {Bamba},
  {Hiraga}, {Senda}, {Katayama}, {Kitamoto}, {Tsujimoto}, {Kohmura}, {Tsuboi},
  \& {Awaki}}]{koyama2007}
{Koyama}, K., {Tsunemi}, H., {Dotani}, T., {et~al.} 2007, \pasj, 59, S23

\bibitem[{{Kuulkers} {et~al.}(2008){Kuulkers}, {Ibarra}, {Page}, {Beardmore},
  {Evans}, {Osborne}, {Bode}, {Drake}, {Mukai}, {Ness}, {Orio}, {Saxton},
  {Starrfield}, {Eyres}, {O'Brien}, \& {Muxlow}}]{kuulkers2008}
{Kuulkers}, E., {Ibarra}, A., {Page}, K.~L., {et~al.} 2008, The Astronomer's
  Telegram, 1480, 1

\bibitem[{{Liller} \& {Jones}(1999)}]{liller1999}
{Liller}, W. \& {Jones}, A. 1999, Information Bulletin on Variable Stars, 4774,
  1

\bibitem[{{Lloyd} {et~al.}(1992){Lloyd}, {O'Brien}, {Bode}, {Predehl},
  {Schmitt}, {Truemper}, {Watson}, \& {Pounds}}]{lloyd1992}
{Lloyd}, H.~M., {O'Brien}, T.~J., {Bode}, M.~F., {et~al.} 1992, \nat, 356, 222

\bibitem[{{Lynch} {et~al.}(2008){Lynch}, {Russell}, {Rudy}, {Woodward}, \&
  {Schwarz}}]{lynch2008a}
{Lynch}, D.~K., {Russell}, R.~W., {Rudy}, R.~J., {Woodward}, C.~E., \&
  {Schwarz}, G.~J. 2008, \iaucirc, 8935, 1

\bibitem[{{Mason} {et~al.}(2001){Mason}, {Breeveld}, {Much}, {Carter},
  {Cordova}, {Cropper}, {Fordham}, {Huckle}, {Ho}, {Kawakami}, {Kennea},
  {Kennedy}, {Mittaz}, {Pandel}, {Priedhorsky}, {Sasseen}, {Shirey}, {Smith},
  \& {Vreux}}]{mason2001}
{Mason}, K.~O., {Breeveld}, A., {Much}, R., {et~al.} 2001, \aap, 365, L36

\bibitem[{{Maughan} {et~al.}(2004){Maughan}, {Jones}, {Ebeling}, \&
  {Scharf}}]{maughan2004}
{Maughan}, B.~J., {Jones}, L.~R., {Ebeling}, H., \& {Scharf}, C. 2004, \mnras,
  351, 1193

\bibitem[{{Mitsuda} {et~al.}(2007){Mitsuda}, {Bautz}, {Inoue}, {Kelley},
  {Koyama}, {Kunieda}, {Makishima}, {Ogawara}, {Petre}, {Takahashi}, {Tsunemi},
  {White}, {Anabuki}, {Angelini}, {Arnaud}, {Awaki}, {Bamba}, {Boyce}, {Brown},
  {Chan}, {Cottam}, {Dotani}, {Doty}, {Ebisawa}, {Ezoe}, {Fabian}, {Figueroa},
  {Fujimoto}, {Fukazawa}, {Furusho}, {Furuzawa}, {Gendreau}, {Griffiths},
  {Haba}, {Hamaguchi}, {Harrus}, {Hasinger}, {Hatsukade}, {Hayashida}, {Henry},
  {Hiraga}, {Holt}, {Hornschemeier}, {Hughes}, {Hwang}, {Ishida}, {Ishisaki},
  {Isobe}, {Itoh}, {Iyomoto}, {Kahn}, {Kamae}, {Katagiri}, {Kataoka},
  {Katayama}, {Kawai}, {Kilbourne}, {Kinugasa}, {Kissel}, {Kitamoto}, {Kohama},
  {Kohmura}, {Kokubun}, {Kotani}, {Kotoku}, {Kubota}, {Madejski}, {Maeda},
  {Makino}, {Markowitz}, {Matsumoto}, {Matsumoto}, {Matsuoka}, {Matsushita},
  {McCammon}, {Mihara}, {Misaki}, {Miyata}, {Mizuno}, {Mori}, {Mori}, {Morii},
  {Moseley}, {Mukai}, {Murakami}, {Murakami}, {Mushotzky}, {Nagase}, {Namiki},
  {Negoro}, {Nakazawa}, {Nousek}, {Okajima}, {Ogasaka}, {Ohashi}, {Oshima},
  {Ota}, {Ozaki}, {Ozawa}, {Parmar}, {Pence}, {Porter}, {Reeves}, {Ricker},
  {Sakurai}, {Sanders}, {Senda}, {Serlemitsos}, {Shibata}, {Soong}, {Smith},
  {Suzuki}, {Szymkowiak}, {Takahashi}, {Tamagawa}, {Tamura}, {Tamura},
  {Tanaka}, {Tashiro}, {Tawara}, {Terada}, {Terashima}, {Tomida}, {Torii},
  {Tsuboi}, {Tsujimoto}, {Tsuru}, {Turner}, {Ueda}, {Ueno}, {Ueno}, {Uno},
  {Urata}, {Watanabe}, {Yamamoto}, {Yamaoka}, {Yamasaki}, {Yamashita},
  {Yamauchi}, {Yamauchi}, {Yaqoob}, {Yonetoku}, \& {Yoshida}}]{mitsuda2007}
{Mitsuda}, K., {Bautz}, M., {Inoue}, H., {et~al.} 2007, \pasj, 59, S1

\bibitem[{{Mukai} \& {Ishida}(2001)}]{mukai2001}
{Mukai}, K. \& {Ishida}, M. 2001, \apj, 551, 1024

\bibitem[{{Mukai} \& {Orio}(2005)}]{mukai2005}
{Mukai}, K. \& {Orio}, M. 2005, \apj, 622, 602

\bibitem[{{Mukai} \& {Shiokawa}(1993)}]{mukai1993}
{Mukai}, K. \& {Shiokawa}, K. 1993, \apj, 418, 863

\bibitem[{{Mukerjee} {et~al.}(2004){Mukerjee}, {Osborne}, {Wells}, {Abbey},
  {Beardmore}, {Short}, {Ambrosi}, \& {Moretti}}]{mukerjee2004}
{Mukerjee}, K., {Osborne}, J.~P., {Wells}, A.~A., {et~al.} 2004, in Society of
  Photo-Optical Instrumentation Engineers (SPIE) Conference Series, Vol. 5165,
  Society of Photo-Optical Instrumentation Engineers (SPIE) Conference Series,
  ed. {K.~A.~Flanagan \& O.~H.~W.~Siegmund}, 251--261

\bibitem[{{Nakajima} {et~al.}(2008){Nakajima}, {Yamaguchi}, {Matsumoto},
  {Tsuru}, {Koyama}, {Tsunemi}, {Hayashida}, {Torii}, {Namiki}, {Katsuda},
  {Shoji}, {Matsuura}, {Miyauchi}, {Dotani}, {Ozaki}, {Murakami}, {Bautz},
  {Kissel}, {Lamarr}, \& {Prigozhin}}]{nakajima2008}
{Nakajima}, H., {Yamaguchi}, H., {Matsumoto}, H., {et~al.} 2008, \pasj, 60, 1

\bibitem[{{Nakano} {et~al.}(2008){Nakano}, {Beize}, {Jin}, {Gao}, {Yamaoka},
  {Haseda}, {Guido}, {Sostero}, {Klingenberg}, \& {Kadota}}]{nakano2008}
{Nakano}, S., {Beize}, J., {Jin}, Z., {et~al.} 2008, \iaucirc, 8934, 1

\bibitem[{{Nelson} {et~al.}(2008){Nelson}, {Orio}, {Cassinelli}, {Still},
  {Leibowitz}, \& {Mucciarelli}}]{nelson2008}
{Nelson}, T., {Orio}, M., {Cassinelli}, J.~P., {et~al.} 2008, \apj, 673, 1067

\bibitem[{{Ness}(2010)}]{ness2010a}
{Ness}, J.-U. 2010, Astronomische Nachrichten, 331, 179

\bibitem[{{Ness} {et~al.}(2009){Ness}, {Drake}, {Beardmore}, {Boyd}, {Bode},
  {Brady}, {Evans}, {Gaensicke}, {Kitamoto}, {Knigge}, {Miller}, {Osborne},
  {Page}, {Rodriguez-Gil}, {Schwarz}, {Staels}, {Steeghs}, {Takei},
  {Tsujimoto}, {Wesson}, \& {Zijlstra}}]{ness2009b}
{Ness}, J.-U., {Drake}, J.~J., {Beardmore}, A.~P., {et~al.} 2009, \aj, 137,
  4160

\bibitem[{{Ness} {et~al.}(2011){Ness}, {Osborne}, {Dobrotka}, {Page}, {Drake},
  {Pinto}, {Detmers}, {Schwarz}, {Bode}, \& {Beardmore}}]{ness2011}
{Ness}, J.-U., {Osborne}, J.~P., {Dobrotka}, A., {et~al.} 2011, \mnras,
  submitted.

\bibitem[{{Ness} {et~al.}(2008{\natexlab{a}}){Ness}, {Starrfield}, {Gonzalez},
  {Kuulkers}, {Osborne}, {Page}, {Schwarz}, {Vanlandingham}, {Drake},
  {Hernanz}, {Evans}, {Gehrels}, {Krautter}, {Gehrz}, \&
  {Woodward}}]{ness2008a}
{Ness}, J.-U., {Starrfield}, S., {Gonzalez}, R., {et~al.} 2008{\natexlab{a}},
  The Astronomer's Telegram, 1561, 1

\bibitem[{{Ness} {et~al.}(2008{\natexlab{b}}){Ness}, {Starrfield}, {Gonzalez},
  {Kuulkers}, {Osborne}, {Page}, {Schwarz}, {Vanlandingham}, {Drake},
  {Hernanz}, {Sala}, {Evans}, {Gehrels}, {Hauschildt}, {Krautter}, {Gehrz}, \&
  {Woodward}}]{ness2008b}
{Ness}, J.-U., {Starrfield}, S., {Gonzalez}, R., {et~al.} 2008{\natexlab{b}},
  The Astronomer's Telegram, 1573, 1

\bibitem[{{Norton} \& {Mukai}(2007)}]{norton2007}
{Norton}, A.~J. \& {Mukai}, K. 2007, \aap, 472, 225

\bibitem[{{Norton} \& {Watson}(1989)}]{norton1989}
{Norton}, A.~J. \& {Watson}, M.~G. 1989, \mnras, 237, 715

\bibitem[{{Orio} {et~al.}(2009){Orio}, {Mukai}, {Bianchini}, {de Martino}, \&
  {Howell}}]{orio2009}
{Orio}, M., {Mukai}, K., {Bianchini}, A., {de Martino}, D., \& {Howell}, S.
  2009, \apj, 690, 1753

\bibitem[{{Orio} {et~al.}(2001){Orio}, {Parmar}, {Benjamin}, {Amati},
  {Frontera}, {Greiner}, {{\"O}gelman}, {Mineo}, {Starrfield}, \&
  {Trussoni}}]{orio2001}
{Orio}, M., {Parmar}, A., {Benjamin}, R., {et~al.} 2001, \mnras, 326, L13

\bibitem[{{Osborne} {et~al.}(2008){Osborne}, {Page}, {Evans}, {Beardmore},
  {Kuulkers}, {Ibarra}, {Ness}, {Starrfield}, {Bode}, {Drake}, {Orio},
  {Krautter}, {Tsujimoto}, \& {Takei}}]{osborne2008}
{Osborne}, J.~P., {Page}, K., {Evans}, P., {et~al.} 2008, The Astronomer's
  Telegram, 1542, 1

\bibitem[{{Osborne} {et~al.}(2011){Osborne}, {Page}, {Beardmore}, {Bode},
  {Goad}, {O'Brien}, {Starrfield}, {Rauch}, {Ness}, {Krautter}, {Schwarz},
  {Burrows}, {Gehrels}, {Drake}, {Evans}, \& {Eyres}}]{osborne2011}
{Osborne}, J.~P., {Page}, K.~L., {Beardmore}, A.~P., {et~al.} 2011, \apj,
  submitted.

\bibitem[{{Page} {et~al.}(2008){Page}, {Osborne}, {Evans}, {Kuulkers},
  {Ibarra}, {Drake}, {Ness}, {Beardmore}, {Bode}, {Orio}, \&
  {Schwarz}}]{page2008}
{Page}, K.~L., {Osborne}, J.~P., {Evans}, P.~A., {et~al.} 2008, The
  Astronomer's Telegram, 1523, 1

\bibitem[{{Page} {et~al.}(2010){Page}, {Osborne}, {Evans}, {Wynn}, {Beardmore},
  {Starling}, {Bode}, {Ibarra}, {Kuulkers}, {Ness}, \& {Schwarz}}]{page2010}
{Page}, K.~L., {Osborne}, J.~P., {Evans}, P.~A., {et~al.} 2010, \mnras, 401,
  121

\bibitem[{{Page} {et~al.}(2009){Page}, {Osborne}, {Read}, {Evans}, {Ness},
  {Beardmore}, {Bode}, {Schwarz}, \& {Starrfield}}]{page2009}
{Page}, K.~L., {Osborne}, J.~P., {Read}, A.~M., {et~al.} 2009, \aap, 507, 923

\bibitem[{{Pagnotta} {et~al.}(2009){Pagnotta}, {Schaefer}, {Xiao}, {Collazzi},
  \& {Kroll}}]{pagnotta2009}
{Pagnotta}, A., {Schaefer}, B.~E., {Xiao}, L., {Collazzi}, A.~C., \& {Kroll},
  P. 2009, \aj, 138, 1230

\bibitem[{{Patterson}(1994)}]{patterson1994}
{Patterson}, J. 1994, \pasp, 106, 209

\bibitem[{{Payne-Gaposchkin}(1957)}]{payne1957}
{Payne-Gaposchkin}, C. 1957, {The Galactic Novae}, ed. C.~{Payne-Gaposchkin}

\bibitem[{{Prialnik}(1986)}]{prialnik1986a}
{Prialnik}, D. 1986, \apj, 310, 222

\bibitem[{{Prialnik} \& {Kovetz}(1995)}]{prialnik1995}
{Prialnik}, D. \& {Kovetz}, A. 1995, \apj, 445, 789

\bibitem[{{Prialnik} {et~al.}(1982){Prialnik}, {Livio}, {Shaviv}, \&
  {Kovetz}}]{prialnik1982}
{Prialnik}, D., {Livio}, M., {Shaviv}, G., \& {Kovetz}, A. 1982, \apj, 257, 312

\bibitem[{{Prialnik} \& {Shara}(1986)}]{prialnik1986b}
{Prialnik}, D. \& {Shara}, M.~M. 1986, \apj, 311, 172

\bibitem[{{Ribeiro} \& {Diaz}(2008)}]{ribeiro2008}
{Ribeiro}, F.~M.~A. \& {Diaz}, M.~P. 2008, \pasj, 60, 327

\bibitem[{{Roming} {et~al.}(2005){Roming}, {Kennedy}, {Mason}, {Nousek}, {Ahr},
  {Bingham}, {Broos}, {Carter}, {Hancock}, {Huckle}, {Hunsberger}, {Kawakami},
  {Killough}, {Koch}, {McLelland}, {Smith}, {Smith}, {Soto}, {Boyd},
  {Breeveld}, {Holland}, {Ivanushkina}, {Pryzby}, {Still}, \&
  {Stock}}]{roming2005}
{Roming}, P.~W.~A., {Kennedy}, T.~E., {Mason}, K.~O., {et~al.} 2005, Space
  Science Reviews, 120, 95

\bibitem[{{Rothenflug} \& {Arnaud}(1985)}]{rothenflug1985}
{Rothenflug}, R. \& {Arnaud}, M. 1985, \aap, 144, 431

\bibitem[{{Rudy} {et~al.}(2008){Rudy}, {Lynch}, {Russell}, {Woodward}, \&
  {Covey}}]{rudy2008}
{Rudy}, R.~J., {Lynch}, D.~K., {Russell}, R.~W., {Woodward}, C.~E., \& {Covey},
  K. 2008, \iaucirc, 8938, 2

\bibitem[{{Samus}(2008)}]{samus2008}
{Samus}, N.~N. 2008, \iaucirc, 8934, 2

\bibitem[{{Saxton} {et~al.}(2008){Saxton}, {Read}, {Esquej}, {Freyberg},
  {Altieri}, \& {Bermejo}}]{saxton2008}
{Saxton}, R.~D., {Read}, A.~M., {Esquej}, P., {et~al.} 2008, VizieR Online Data
  Catalog, 348, 611

\bibitem[{{Schaefer}(2010)}]{schaefer2010a}
{Schaefer}, B.~E. 2010, \apj, 187, S275

\bibitem[{{Schaefer} \& {Collazzi}(2010)}]{schaefer2010b}
{Schaefer}, B.~E. \& {Collazzi}, A.~C. 2010, \aj, 139, 1831

\bibitem[{{Serlemitsos} {et~al.}(2007){Serlemitsos}, {Soong}, {Chan},
  {Okajima}, {Lehan}, {Maeda}, {Itoh}, {Mori}, {Iizuka}, {Itoh}, {Inoue},
  {Okada}, {Yokoyama}, {Itoh}, {Ebara}, {Nakamura}, {Suzuki}, {Ishida},
  {Hayakawa}, {Inoue}, {Okuma}, {Kubota}, {Suzuki}, {Osawa}, {Yamashita},
  {Kunieda}, {Tawara}, {Ogasaka}, {Furuzawa}, {Tamura}, {Shibata}, {Haba},
  {Naitou}, \& {Misaki}}]{serlemitsos2007}
{Serlemitsos}, P.~J., {Soong}, Y., {Chan}, K., {et~al.} 2007, \pasj, 59, S9

\bibitem[{{Shara} {et~al.}(1986){Shara}, {Livio}, {Moffat}, \&
  {Orio}}]{shara1986}
{Shara}, M.~M., {Livio}, M., {Moffat}, A.~F.~J., \& {Orio}, M. 1986, \apj, 311,
  163

\bibitem[{{Smith} {et~al.}(2001){Smith}, {Brickhouse}, {Liedahl}, \&
  {Raymond}}]{smith2001}
{Smith}, R.~K., {Brickhouse}, N.~S., {Liedahl}, D.~A., \& {Raymond}, J.~C.
  2001, \apj, 556, L91

\bibitem[{{Sokoloski} {et~al.}(2006){Sokoloski}, {Luna}, {Mukai}, \&
  {Kenyon}}]{sokoloski2006}
{Sokoloski}, J.~L., {Luna}, G.~J.~M., {Mukai}, K., \& {Kenyon}, S.~J. 2006,
  \nat, 442, 276

\bibitem[{{Starrfield} {et~al.}(2008){Starrfield}, {Illiadis}, \&
  {Hix}}]{starrfield2008}
{Starrfield}, J.~H., {Illiadis}, C., \& {Hix}, W.~R. 2008, {in Classical
  Novae}, ed. {M.~Bode and A.~Evans} ({Cambridge: Cambridge University Press}),
  77--101

\bibitem[{{Str{\"u}der} {et~al.}(2001){Str{\"u}der}, {Briel}, {Dennerl},
  {Hartmann}, {Kendziorra}, {Meidinger}, {Pfeffermann}, {Reppin}, {Aschenbach},
  {Bornemann}, {Br{\"a}uninger}, {Burkert}, {Elender}, {Freyberg}, {Haberl},
  {Hartner}, {Heuschmann}, {Hippmann}, {Kastelic}, {Kemmer}, {Kettenring},
  {Kink}, {Krause}, {M{\"u}ller}, {Oppitz}, {Pietsch}, {Popp}, {Predehl},
  {Read}, {Stephan}, {St{\"o}tter}, {Tr{\"u}mper}, {Holl}, {Kemmer}, {Soltau},
  {St{\"o}tter}, {Weber}, {Weichert}, {von Zanthier}, {Carathanassis}, {Lutz},
  {Richter}, {Solc}, {B{\"o}ttcher}, {Kuster}, {Staubert}, {Abbey}, {Holland},
  {Turner}, {Balasini}, {Bignami}, {La Palombara}, {Villa}, {Buttler},
  {Gianini}, {Lain{\'e}}, {Lumb}, \& {Dhez}}]{struder2001}
{Str{\"u}der}, L., {Briel}, U., {Dennerl}, K., {et~al.} 2001, \aap, 365, L18

\bibitem[{{Takahashi} {et~al.}(2007){Takahashi}, {Abe}, {Endo}, {Endo}, {Ezoe},
  {Fukazawa}, {Hamaya}, {Hirakuri}, {Hong}, {Horii}, {Inoue}, {Isobe}, {Itoh},
  {Iyomoto}, {Kamae}, {Kasama}, {Kataoka}, {Kato}, {Kawaharada}, {Kawano},
  {Kawashima}, {Kawasoe}, {Kishishita}, {Kitaguchi}, {Kobayashi}, {Kokubun},
  {Kotoku}, {Kouda}, {Kubota}, {Kuroda}, {Madejski}, {Makishima}, {Masukawa},
  {Matsumoto}, {Mitani}, {Miyawaki}, {Mizuno}, {Mori}, {Mori}, {Murashima},
  {Murakami}, {Nakazawa}, {Niko}, {Nomachi}, {Okada}, {Ohno}, {Oonuki}, {Ota},
  {Ozawa}, {Sato}, {Shinoda}, {Sugiho}, {Suzuki}, {Taguchi}, {Takahashi},
  {Takahashi}, {Takeda}, {Tamura}, {Tamura}, {Tanaka}, {Tanihata}, {Tashiro},
  {Terada}, {Tominaga}, {Uchiyama}, {Watanabe}, {Yamaoka}, {Yanagida}, \&
  {Yonetoku}}]{takahashi2007}
{Takahashi}, T., {Abe}, K., {Endo}, M., {et~al.} 2007, \pasj, 59, S35

\bibitem[{{Takei} {et~al.}(2009){Takei}, {Tsujimoto}, {Kitamoto}, {Ness},
  {Drake}, {Takahashi}, \& {Mukai}}]{takei2009}
{Takei}, D., {Tsujimoto}, M., {Kitamoto}, S., {et~al.} 2009, \apj, 697, L54

\bibitem[{{Talavera}(2009)}]{telavera2009}
{Talavera}, A. 2009, \apss, 320, 177

\bibitem[{{Tomov} {et~al.}(2008{\natexlab{a}}){Tomov}, {Mikolajewski},
  {Brozek}, {Ragan}, {Swierczynski}, {Wychudzki}, \& {Galan}}]{tomov2008b}
{Tomov}, T., {Mikolajewski}, M., {Brozek}, T., {et~al.} 2008{\natexlab{a}}, The
  Astronomer's Telegram, 1485, 1

\bibitem[{{Tomov} {et~al.}(2008{\natexlab{b}}){Tomov}, {Mikolajewski}, {Ragan},
  {Swierczynski}, \& {Wychudzki}}]{tomov2008a}
{Tomov}, T., {Mikolajewski}, M., {Ragan}, E., {Swierczynski}, E., \&
  {Wychudzki}, P. 2008{\natexlab{b}}, The Astronomer's Telegram, 1475, 1

\bibitem[{{Tsujimoto} {et~al.}(2005){Tsujimoto}, {Feigelson}, {Grosso},
  {Micela}, {Tsuboi}, {Favata}, {Shang}, \& {Kastner}}]{tsujimoto2005}
{Tsujimoto}, M., {Feigelson}, E.~D., {Grosso}, N., {et~al.} 2005, \apj, 160,
  S503

\bibitem[{{Tsujimoto} {et~al.}(2009){Tsujimoto}, {Takei}, {Drake}, {Ness}, \&
  {Kitamoto}}]{tsujimoto2009}
{Tsujimoto}, M., {Takei}, D., {Drake}, J.~J., {Ness}, J.-U., \& {Kitamoto}, S.
  2009, \pasj, 61, S69

\bibitem[{{Turner} {et~al.}(2001){Turner}, {Abbey}, {Arnaud}, {Balasini},
  {Barbera}, {Belsole}, {Bennie}, {Bernard}, {Bignami}, {Boer}, {Briel},
  {Butler}, {Cara}, {Chabaud}, {Cole}, {Collura}, {Conte}, {Cros}, {Denby},
  {Dhez}, {Di Coco}, {Dowson}, {Ferrando}, {Ghizzardi}, {Gianotti}, {Goodall},
  {Gretton}, {Griffiths}, {Hainaut}, {Hochedez}, {Holland}, {Jourdain},
  {Kendziorra}, {Lagostina}, {Laine}, {La Palombara}, {Lortholary}, {Lumb},
  {Marty}, {Molendi}, {Pigot}, {Poindron}, {Pounds}, {Reeves}, {Reppin},
  {Rothenflug}, {Salvetat}, {Sauvageot}, {Schmitt}, {Sembay}, {Short},
  {Spragg}, {Stephen}, {Str{\"u}der}, {Tiengo}, {Trifoglio}, {Tr{\"u}mper},
  {Vercellone}, {Vigroux}, {Villa}, {Ward}, {Whitehead}, \&
  {Zonca}}]{turner2001}
{Turner}, M.~J.~L., {Abbey}, A., {Arnaud}, M., {et~al.} 2001, \aap, 365, L27

\bibitem[{{Voges} {et~al.}(2000){Voges}, {Aschenbach}, {Boller}, {Brauninger},
  {Briel}, {Burkert}, {Dennerl}, {Englhauser}, {Gruber}, {Haberl}, {Hartner},
  {Hasinger}, {Pfeffermann}, {Pietsch}, {Predehl}, {Schmitt}, {Trumper}, \&
  {Zimmermann}}]{voges2000}
{Voges}, W., {Aschenbach}, B., {Boller}, T., {et~al.} 2000, VizieR Online Data
  Catalog, 9029, 0

\bibitem[{{Warner}(2003)}]{warner2003}
{Warner}, B. 2003, {Cataclysmic Variable Stars}, ed. {Warner, B.} ({Cambridge:
  Cambridge University Press})

\bibitem[{{Warner}(2008)}]{warner2008}
{Warner}, B. 2008, {in Classical Novae}, ed. {Bode, M.~F.~\& Evans, A.}
  ({Cambridge: Cambridge University Press}), 16--33

\bibitem[{{Wilms} {et~al.}(2000){Wilms}, {Allen}, \& {McCray}}]{wilms2000}
{Wilms}, J., {Allen}, A., \& {McCray}, R. 2000, \apj, 542, 914

\end{thebibliography}

\end{document}